\documentclass[12pt,a4paper]{article}
\usepackage{amsmath,amssymb}
\newcommand{\beq}{\begin{eqnarray}}
\newcommand{\eeq}{\end{eqnarray}}

\newcommand{\be}{\begin{equation}}
\newcommand{\ee}{\end{equation}}

\newcommand{\bm}{\begin{multline}}
\newcommand{\fm}{\end{multline}}

\begin{document}
\setlength{\unitlength}{.8mm}

\begin{titlepage} 
\vspace*{0.5cm}
\begin{center}
{\Large\bf Nonlinear integral equations for the finite size effects of RSOS and vertex-models
and related quantum field theories}
\end{center}
\vspace{0.2cm}
\begin{center}
{\large \'Arp\'ad Heged\H us}
\end{center}
\bigskip

\begin{center}
Istituto Nazionale di Fisica Nucleare, Sezione di Bologna, \\
       Via Irnerio 46, 40126 Bologna, Italy
\end{center}

\vspace{0.1cm}

\begin{center}
and
\end{center}

\vspace{0.1cm}

\begin{center}
Research Institute for Particle and Nuclear Physics,\\
Hungarian Academy of Sciences,\\
H-1525 Budapest 114, P.O.B. 49, Hungary\\ 
\end{center}
%
\begin{abstract}
Starting from critical RSOS lattice models with appropriate inhomogeneities,
we derive two component nonlinear integral equations to describe the
finite volume ground state
energy of the massive $\phi_{\mbox{id,id,adj}}$ perturbation of the $SU(2)_k \times SU(2)_{k'}
/SU(2)_{k+k'}$ coset models. 
When $k' \to \infty$ while the value of $k$ is fixed, the equations
correspond to the current-current perturbation of the $SU(2)_k$ WZW model.
Then modifying one of the kernel functions of these equations, we propose two component
nonlinear integral equations for the fractional supersymmetric sine-Gordon models.
 The lattice versions of our equations describe the finite size effects in the corresponding lattice models, 
namely in the critical RSOS($k,q$) models, in the isotropic higher-spin vertex models, and 
in the anisotropic higher-spin vertex models. Numerical and analytical checks are also
performed to confirm the correctness of our equations. These type of equations
make it easier to treat the excited state problem.
\end{abstract}

\end{titlepage}


\newsavebox{\SSddv}
\sbox{\SSddv}{\begin{picture}(140,25) (-70,-12.5)
\setlength{\unitlength}{.5mm} 
\put(0,0){\circle*{3}}
\put(10,0){\circle{3}}
\put(20,0){\circle{3}}
\put(30,0){\circle{3}}
\put(40,10){\circle{3}}
\put(40,-10){\circle{3}}
\put(-10,0){\circle{3}}
\put(-20,0){\circle{3}}
\put(-30,0){\circle{3}}
\put(-40,10){\circle{3}}
\put(-40,-10){\circle{3}}

\put(1.5,0){\line(1,0){7}}
\put(21.5,0){\line(1,0){7}}
\put(-8.5,0){\line(1,0){7}}
\put(-28.5,0){\line(1,0){7}}

\put(31.1,1.1){\line(1,1){7.7}}
\put(31.1,-1.1){\line(1,-1){7.7}}
\put(-31.1,1.1){\line(-1,1){7.7}}
\put(-31.1,-1.1){\line(-1,-1){7.7}}

\put(58,-2){$a$}

\multiput(12.5,0) (1,0) {6} {\circle*{0.2}}
\multiput(-17.5,0) (1,0) {6} {\circle*{0.2}}
\setlength{\unitlength}{1mm} 

\thicklines{
\put(13.5,0){\oval(28,15)[]}
\put(-15.8,0){\oval(23,15)[]}
}
\setlength{\unitlength}{.5mm} 
\put(2,-3){\makebox(0,0)[t]{{\protect\tiny 0}}}
\put(12,-3){\makebox(0,0)[t]{{\protect\tiny 1}}}

\put(29,-3){\makebox(0,0)[t]{{\protect\tiny  $p$--2}}}
\put(44,-10.5){\makebox(0,0)[t]{{\protect\tiny $p$--1}}}
\put(44,11){\makebox(0,0)[t]{{\protect\tiny $p$}}}
\put(-12,-3){\makebox(0,0)[t]{{\protect\tiny --1}}}

\put(-28,-3){\makebox(0,0)[t]{{\protect\tiny 2--$\tilde{p}$}}}
\put(-45,-10){\makebox(0,0)[t]{{\protect\tiny 1--$\tilde{p}$}}}
\put(-45,11){\makebox(0,0)[t]{{\protect\tiny --$\tilde{p}$}}}
\put(28,-17){\makebox(0,0)[t]{{\protect\tiny $a_0(x),\bar{a}_0(x),G_p(x)$}}}
\put(-29,-17){\makebox(0,0)[t]{{\protect\tiny $a(x),\bar{a}(x),G_{\tilde{p}-1}(x)$}}}

\end{picture}}



\newsavebox{\FSSGN}
\sbox{\FSSGN}{\begin{picture}(140,25) (-70,-12.5)
\setlength{\unitlength}{.5mm} 
\put(0,0){\circle*{3}}
\put(10,0){\circle{3}}
\put(20,0){\circle{3}}
\put(30,0){\circle{3}}
\put(40,10){\circle{3}}
\put(40,-10){\circle{3}}
\put(-10,0){\circle{3}}
\put(-20,0){\circle{3}}
\put(-30,0){\circle{3}}
\put(-40,0){\circle{3}}

\put(1.5,0){\line(1,0){7}}
\put(21.5,0){\line(1,0){7}}
\put(-8.5,0){\line(1,0){7}}
\put(-28.5,0){\line(1,0){7}}
\put(-38.5,0){\line(1,0){7}}

\put(31.1,1.1){\line(1,1){7.7}}
\put(31.1,-1.1){\line(1,-1){7.7}}

\put(58,-2){$b$}

\multiput(12.5,0) (1,0) {6} {\circle*{0.2}}
\multiput(-17.5,0) (1,0) {6} {\circle*{0.2}}
\setlength{\unitlength}{1mm} 

\thicklines{
\put(13.5,0){\oval(28,15)[]}
\put(-13.2,0){\oval(18,12)[]}
}
\setlength{\unitlength}{.5mm} 
\put(2,-3){\makebox(0,0)[t]{{\protect\tiny $k$}}}
\put(12,-3){\makebox(0,0)[t]{{\protect\tiny $k$+1}}}

\put(29,-3){\makebox(0,0)[t]{{\protect\tiny  $r$--2}}}
\put(44,-11.0){\makebox(0,0)[t]{{\protect\tiny $r$--1}}}
\put(44,11){\makebox(0,0)[t]{{\protect\tiny $r$}}}
\put(-14,-3){\makebox(0,0)[t]{{\protect\tiny $k$--1}}}

\put(-27,-3){\makebox(0,0)[t]{{\protect\tiny 2}}}
\put(-38,-3){\makebox(0,0)[t]{{\protect\tiny 1}}}

\put(28,-17){\makebox(0,0)[t]{{\protect\tiny $a_0(x),\bar{a}_0(x),G(x)$}}}
\put(-26,-14.5){\makebox(0,0)[t]{{\protect\tiny $a(x),\bar{a}(x),\tilde{G}(x)$}}}

\end{picture}}



\newsavebox{\FSSGT}
\sbox{\FSSGT}{\begin{picture}(140,25) (-70,-12.5)
\setlength{\unitlength}{0.6mm} 
\put(0,0){\circle*{3}}
\put(10,0){\circle{3}}
\put(20,0){\circle{3}}
\put(30,0){\circle{3}}
\put(40,10){\circle{3}}
\put(40,-10){\circle{3}}
\put(-10,0){\circle{3}}
\put(-20,0){\circle{3}}
\put(-30,0){\circle{3}}
\put(-40,0){\circle{3}}

\put(1.5,0){\line(1,0){7}}
\put(21.5,0){\line(1,0){7}}
\put(-8.5,0){\line(1,0){7}}
\put(-28.5,0){\line(1,0){7}}
\put(-38.5,0){\line(1,0){7}}

\put(31.1,1.1){\line(1,1){7.7}}
\put(31.1,-1.1){\line(1,-1){7.7}}

\put(58,-2){$b$}

\multiput(12.5,0) (1,0) {6} {\circle*{0.2}}
\multiput(-17.5,0) (1,0) {6} {\circle*{0.2}}
\setlength{\unitlength}{1mm} 

\setlength{\unitlength}{.6mm} 
\put(1.5,-3){\makebox(0,0)[t]{{\protect\tiny $k$}}}
\put(11.5,-3){\makebox(0,0)[t]{{\protect\tiny $k$+1}}}

\put(29,-3){\makebox(0,0)[t]{{\protect\tiny  $r$--2}}}
\put(44,-11.0){\makebox(0,0)[t]{{\protect\tiny $r$--1}}}
\put(44,11){\makebox(0,0)[t]{{\protect\tiny $r$}}}
\put(-10.5,-3){\makebox(0,0)[t]{{\protect\tiny $k$--1}}}

\put(-30,-3){\makebox(0,0)[t]{{\protect\tiny 2}}}
\put(-40.2,-3){\makebox(0,0)[t]{{\protect\tiny 1}}}

\end{picture}}


\newsavebox{\suzen}
\sbox{\suzen}{\begin{picture}(140,25) (-70,-12.5)

\put(0,0){\circle*{3}}
\put(10,0){\circle{3}}
\put(20,0){\circle{3}}
\put(-10,0){\circle{3}}
\put(-20,0){\circle{3}}
\put(-30,0){\circle{3}}

\put(1.5,0){\line(1,0){7}}
\put(11.5,0){\line(1,0){7}}
\put(-8.5,0){\line(1,0){7}}
\put(-18.5,0){\line(1,0){7}}
\put(-28.5,0){\line(1,0){7}}

\put(38,-2){$a$}


\multiput(22.5,0) (1,0) {6} {\circle*{0.2}}
\multiput(-37.5,0) (1,0) {6} {\circle*{0.2}}

\thicklines{
\put(17,0){\oval(36.5,15)[]}
\put(-24,0){\oval(31,15)[]}
}
\put(1.5,-3){\makebox(0,0)[t]{{\protect\scriptsize {\em k}}}}
\put(11.7,-3){\makebox(0,0)[t]{{\protect\scriptsize {\em k}+1}}}
\put(21.7,-3){\makebox(0,0)[t]{{\protect\scriptsize{\em k}+2}}}
\put(-13.5,-3){\makebox(0,0)[t]{{\protect\scriptsize {\em k}--1}}}
\put(-21.5,-3){\makebox(0,0)[t]{{\protect\scriptsize {\em k}--2 }}}
\put(-30,-3){\makebox(0,0)[t]{{\protect\scriptsize {\em k}--3 }}}
\put(21,-10){\makebox(0,0)[t]{{\protect\scriptsize  $a_0(x),\bar{a}_0(x),F(x)$ }}}
\put(-21,-10){\makebox(0,0)[t]{{\protect\scriptsize  $a(x),\bar{a}(x),F(x)$ }}}

\end{picture}}


\newsavebox{\RSN}
\sbox{\RSN}{\begin{picture}(140,25) (-70,-12.5)

\put(0,0){\circle*{3}}
\put(10,0){\circle{3}}
\put(20,0){\circle{3}}
\put(30,0){\circle{3}}
\put(-10,0){\circle{3}}
\put(-20,0){\circle{3}}
\put(-30,0){\circle{3}}

\put(1.5,0){\line(1,0){7}}
\put(21.5,0){\line(1,0){7}}
\put(-8.5,0){\line(1,0){7}}
\put(-28.5,0){\line(1,0){7}}

\put(38,-2){$a$}


\multiput(12.5,0) (1,0) {6} {\circle*{0.2}}
\multiput(-17.5,0) (1,0) {6} {\circle*{0.2}}

\thicklines{
\put(17,0){\oval(36.5,15)[]}
\put(-21.7,0){\oval(27,15)[]}
}
\put(1.5,-3){\makebox(0,0)[t]{{\protect\scriptsize {\em k}}}}
\put(11.7,-3){\makebox(0,0)[t]{{\protect\scriptsize {\em k}+1}}}
\put(21.7,-3){\makebox(0,0)[t]{{\protect\scriptsize{\em r}-4}}}
\put(31.0,-3){\makebox(0,0)[t]{{\protect\scriptsize{\em r}-3}}}
\put(-13.5,-3){\makebox(0,0)[t]{{\protect\scriptsize {\em k}--1}}}
\put(-21.5,-3){\makebox(0,0)[t]{{\protect\scriptsize 2 }}}
\put(-30,-3){\makebox(0,0)[t]{{\protect\scriptsize 1 }}}
\put(21,-9){\makebox(0,0)[t]{{\protect\scriptsize  $a_0(x),\bar{a}_0(x),G(x)$ }}}
\put(-20,-9){\makebox(0,0)[t]{{\protect\scriptsize  $a(x),\bar{a}(x),\tilde{G}(x)$ }}}

\end{picture}}



\newsavebox{\RSNI}
\sbox{\RSNI}{\begin{picture}(140,25) (-70,-12.5)

\put(0,0){\circle*{3}}
\put(10,0){\circle{3}}
\put(20,0){\circle{3}}
\put(30,0){\circle{3}}
\put(-10,0){\circle{3}}
\put(-20,0){\circle{3}}
\put(-30,0){\circle{3}}

\put(1.5,0){\line(1,0){7}}
\put(21.5,0){\line(1,0){7}}
\put(-8.5,0){\line(1,0){7}}
\put(-28.5,0){\line(1,0){7}}

\put(38,-2){$a$}


\multiput(12.5,0) (1,0) {6} {\circle*{0.2}}
\multiput(-17.5,0) (1,0) {6} {\circle*{0.2}}

\thicklines{
\put(17,0){\oval(36.5,15)[]}
}
\put(1.5,-3){\makebox(0,0)[t]{{\protect\scriptsize {\em k}}}}
\put(11.7,-3){\makebox(0,0)[t]{{\protect\scriptsize {\em k}+1}}}
\put(21.7,-3){\makebox(0,0)[t]{{\protect\scriptsize{\em r}-4}}}
\put(31.0,-3){\makebox(0,0)[t]{{\protect\scriptsize{\em r}-3}}}
\put(-10.0,-3){\makebox(0,0)[t]{{\protect\scriptsize {\em k}--1}}}
\put(-20.5,-3){\makebox(0,0)[t]{{\protect\scriptsize 2 }}}
\put(-30,-3){\makebox(0,0)[t]{{\protect\scriptsize 1 }}}
\put(21,-9){\makebox(0,0)[t]{{\protect\scriptsize  $a_0(x),\bar{a}_0(x),G(x)$ }}}

\put(-18,7){\makebox(0,0)[t]{{\protect\scriptsize RSOS($k,q$) $\longrightarrow$}}}

\end{picture}}



\newsavebox{\RSNII}
\sbox{\RSNII}{\begin{picture}(140,25) (-70,-12.5)

\put(0,0){\circle*{3}}
\put(10,0){\circle{3}}
\put(20,0){\circle{3}}
\put(30,0){\circle{3}}
\put(-10,0){\circle{3}}
\put(-20,0){\circle{3}}
\put(-30,0){\circle{3}}

\put(1.5,0){\line(1,0){7}}
\put(21.5,0){\line(1,0){7}}
\put(-8.5,0){\line(1,0){7}}
\put(-28.5,0){\line(1,0){7}}

\put(38,-2){$b$}


\multiput(12.5,0) (1,0) {6} {\circle*{0.2}}
\multiput(-17.5,0) (1,0) {6} {\circle*{0.2}}

\thicklines{
\put(-21.7,0){\oval(27,15)[]}
}
\put(1.5,-3){\makebox(0,0)[t]{{\protect\scriptsize {\em k'}}}}
\put(11.7,-3){\makebox(0,0)[t]{{\protect\scriptsize {\em k'}--1}}}
\put(21.7,-3){\makebox(0,0)[t]{{\protect\scriptsize 2 }}}
\put(31.0,-3){\makebox(0,0)[t]{{\protect\scriptsize 1 }}}
\put(-13.5,-3){\makebox(0,0)[t]{{\protect\tiny {\em k'}+1}}}
\put(-21.5,-3){\makebox(0,0)[t]{{\protect\tiny {\em r}--4 }}}
\put(-30,-3){\makebox(0,0)[t]{{\protect\tiny {\em r}--3 }}}
\put(-20,-9){\makebox(0,0)[t]{{\protect\scriptsize  $a(x),\bar{a}(x),\tilde{G}(x)$ }}}

\put(18,7){\makebox(0,0)[t]{{\protect\scriptsize  $\longleftarrow$  RSOS($k',q'$)}}}

\end{picture}}



\newsavebox{\RST}
\sbox{\RST}{\begin{picture}(140,25) (-70,-12.5)

\put(0,0){\circle*{3}}
\put(10,0){\circle{3}}
\put(20,0){\circle{3}}
\put(30,0){\circle{3}}
\put(-10,0){\circle{3}}
\put(-20,0){\circle{3}}
\put(-30,0){\circle{3}}

\put(1.5,0){\line(1,0){7}}
\put(21.5,0){\line(1,0){7}}
\put(-8.5,0){\line(1,0){7}}
\put(-28.5,0){\line(1,0){7}}

\put(38,-2){$a$}


\multiput(12.5,0) (1,0) {6} {\circle*{0.2}}
\multiput(-17.5,0) (1,0) {6} {\circle*{0.2}}


\put(0.5,-3){\makebox(0,0)[t]{{\protect\scriptsize {\em k}}}}
\put(10.7,-3){\makebox(0,0)[t]{{\protect\scriptsize {\em k}+1}}}
\put(21.7,-3){\makebox(0,0)[t]{{\protect\scriptsize{\em r}-4}}}
\put(31.0,-3){\makebox(0,0)[t]{{\protect\scriptsize{\em r}-3}}}
\put(-10.0,-3){\makebox(0,0)[t]{{\protect\scriptsize {\em k}--1}}}
\put(-20.5,-3){\makebox(0,0)[t]{{\protect\scriptsize 2 }}}
\put(-30,-3){\makebox(0,0)[t]{{\protect\scriptsize 1 }}}

\end{picture}}



\newsavebox{\NSYM}
\sbox{\NSYM}{\begin{picture}(140,25)(-70,-12.5)
\setlength{\unitlength}{1.0mm}
\put(0,0){\circle*{3}}
\put(10,0){\circle{3}}
\put(20,0){\circle{3}}
\put(30,0){\circle{3}}
\put(-10,0){\circle{3}}
\put(-20,0){\circle{3}}
\put(-30,0){\circle{3}}

\put(1.5,0){\line(1,0){7}}
\put(21.5,0){\line(1,0){7}}
\put(-8.5,0){\line(1,0){7}}
\put(-28.5,0){\line(1,0){7}}



\multiput(12.5,0) (1,0) {6} {\circle*{0.2}}
\multiput(-17.5,0) (1,0) {6} {\circle*{0.2}}

\thicklines{
\put(21.7,0){\oval(27,15)[]}
\put(-21.7,0){\oval(27,15)[]}
}
\put(0.5,-3){\makebox(0,0)[t]{{\protect\scriptsize {\em k}}}}
\put(13.5,-3){\makebox(0,0)[t]{{\protect\scriptsize {\em k}+1}}}
\put(28.5,-3){\makebox(0,0)[t]{{\protect\scriptsize 2{\em k}+2}}}
\put(-13.5,-3){\makebox(0,0)[t]{{\protect\scriptsize {\em k}--1}}}
\put(-21.5,-3){\makebox(0,0)[t]{{\protect\scriptsize 2 }}}
\put(-30,-3){\makebox(0,0)[t]{{\protect\scriptsize 1 }}}
\put(23,-9){\makebox(0,0)[t]{{\protect\scriptsize  $a(x),\bar{a}(x),\tilde{G}(x)$ }}}
\put(-20,-9){\makebox(0,0)[t]{{\protect\scriptsize  $a(x),\bar{a}(x),\tilde{G}(x)$ }}}

\end{picture}}


\section{Introduction}

The massive $\phi_{\mbox{id,id,adj}}$ perturbation of the $SU(2)_k \times SU(2)_{k'}
/SU(2)_{k+k'}$ models belong to an interesting set of integrable quantum field theories (QFT).
In the following we will denote them by $\cal{M}$$A^{+}_{k,k+k'+2}$.
 This family of models includes as particular cases
 some other interesting QFTs like the asymptotically free current-current perturbation of the $SU(2)_k$ WZW model and the
 O(4) nonlinear sigma (NLS) model.
 The spectrum and interactions are governed by two dual fractional symmetries $Q^{(k)}$ and $Q^{(k')}$, one of which
 turns into the fractional supersymmetry for $k=1$ \cite{BL,ABL}. The spectrum of the theory
 consists of a set of massive relativistic kinks of the same mass \cite{BL,ABL}. Their factorised scattering  
 was conjectured in \cite{BL}. The kinks $K_{aa',bb'}$ separate different degenerate vacua, each vacuum being labeled
 by two indices $aa'$. $a$ and $a'$ run independently over the states of nodes of an $A_{k+1}$ and $A_{k'+1}$
 Dynkin diagrams respectively, $a \in \{0,1,...,k\} \  $, $a' \in \{0,1,...,k' \} \  $.  
 The stable kinks $K_{aa',bb'}$ follow the Dynkin diagram adjacency rules $a=b \pm 1$,  $a'=b' \pm 1$
 and the kink-kink scattering matrix appears as the tensor product
 \begin{equation}
 S(\theta)=S^{(k)}_{RSG}(\theta) \otimes S^{(k')}_{RSG}(\theta), \label{1}
 \end{equation}
 where $S^{(k)}_{RSG}(\theta)$ is the scattering matrix of the massive $\phi_{13}$ perturbation of 
 the minimal model $SU(2)_1 \times SU(2)_k/SU(2)_{1+k}$, which is  an integrable restriction of the 
 sine-Gordon model \cite{SM,ALC}.
 
 The fractional supersymmetric sine-Gordon models (FSSG) were first proposed as the models obtained by \lq\lq unrestricting"
 the S-matrices (\ref{1}) describing $\cal{M}$$A^{+}_{k,k+k'+2}$ \cite{ABL}.
Thus the scattering matrix of the FSSG models is of the form
\begin{equation}
 S(\theta)=S^{(k)}_{RSG}(\theta) \otimes S_{SG}^{(\beta)}(\theta), \label{2} 
\end{equation}
 where $S_{SG}^{(\beta)}(\theta)$ is the S-matrix of the sine-Gordon model at coupling $\beta$ \cite{ZZ}.
 The particle content of the FSSG consist of a soliton and an antisoliton of mass $m$ together with a number of 
 breathers of mass
 \begin{equation}
 M_j=2m \sin\left( \frac{\pi j \xi}{2} \right), \qquad j=1,...<1/\xi
 \end{equation}
where
\begin{equation}
 \xi=\frac{k \cdot \beta^{2}/8\pi}{1/k-\beta^2 /8\pi},
 \end{equation}
and there are no breathers in the repulsive regime $\xi<1$.

\begin{figure}[htbp]
\begin{center}
\begin{picture}(280,30)(-140,-15)
\put(-163,-7) {\usebox{\RST}}
\put(-74,-7) {\usebox{\FSSGT}}
\put(-130,-15){\parbox{130mm}{\caption{ \label{1f}\protect {\small
Graphical representation of Y-system of the $a$.) $\cal{M}$$A^{+}_{k,r}$ and $b$.) FSSG model }}}}
\end{picture}
\end{center}
\end{figure}
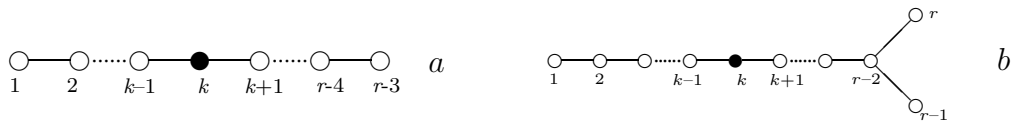

 
 Once the spectrum and the associated S-matrix is known, the TBA method can be employed for calculating the ground state 
energy of the model in finite volume. 
The TBA equations of the $\cal{M}$$A^{+}_{k,r}$ model was conjectured in \cite{FSSS} and the equations can be encoded in an
$A_{r-3}$ Dynkin-diagram (see figure \ref{1f}$a$.)
The TBA equations of the FSSG models were proposed in \cite{fi}, and at the special $\frac{\beta^2}{8\pi}=\frac{r}{r-1}$,
 $1 \leq r \in \mathbb{N}$  values of the coupling, it can be encoded into a ${\mathcal{D}}_{r}$ Dynkin-diagram (see figure \ref{1f}$b$),
 but for general values of the coupling, the TBA integral equations take much complicated form, and the number of unknown
 functions of the equations depend on the continued fraction form of $r$ \cite{TS}.
 The solutions of the TBA equations also satisfy 
the well-known Y-system equations  \cite{ZamiY}.

 In  \cite{fi} it was noted that if a model has an S-matrix in the form of a direct product $S_G \otimes S_H$, and the TBA equations are already
 known for the models described by S-matrices $S_G$ and $S_H$, and the individual TBA equations are encoded on  Dynkin-like 
 diagrams of type $G$ and $H$ respectively, each having one massive node, then the TBA equations for the model with the
 direct product S-matrix can be obtained by gluing the individual TBA equations together at the massive node. This method can 
 also be applied for our perturbed coset model due to the tensor product form of the S-matrix.    
  
  Another approach for calculating finite size effects in a QFT is the use of some integrable lattice regularization
  of the QFT which can be solved by the Bethe Ansatz method.
   In models that can be solved by Bethe Ansatz, the well-known T- and Y-systems  \cite{Kuniba,KP} naturally appear and from them,
   using the analytical properties of the T-functions, one can easily derive the same TBA equations  which 
   can be obtained from the S-matrix.
    Using such an integrable regularization, an alternative method was worked out by Destri and de Vega  \cite{DdV0} for calculating
 finite size effects in the SG model. This is called the nonlinear integral equation (NLIE) technique. The advantage of this
 method is that one gets a single integral equation for any real value of the coupling constant, and the method can be extended
 to excited states, too  \cite{DdV2,DdV1,DdVres}. The problem is that this method can only be applied when the ground state of the model is
 formed by real Bethe Ansatz roots. In most  cases this is not true. For example the ground state of the spin-$S$  $XXX$-chain 
 is formed by $2S$-strings \cite{TB,pcm0,pcm1,KR}.
  
   Recently J. Suzuki \cite{suz} managed to derive some new types of nonlinear integral equations for describing the thermodynamics of
    the higher spin $XXX$-model at finite temperature, which can be regarded as the 
   generalization of the spin-$\frac12$ case which was treated successfully earlier  \cite{DdV0,KBP}. These new equations can be regarded
   as a particular mixture of TBA and NLIE, the special case of which is the 
   Destri-de Vega equation  \cite{DdV0} in the spin-$\frac12 $ case.
    
	 Motivated by the gluing idea  \cite{fi} and Suzuki's results  \cite{suz} Dunning proposed nonlinear integral equations 
       \cite{dun}
	similar to Suzuki's equations for describing the finite size effects in some perturbed conformal field 
	theories whose S-matrices can be written in the form 
	of a direct product, comprising the $\cal{M}$$A^{+}_{k,r}$ models, the fractional 
	supersymmetric sine-Gordon models (FSSG), and the SS-model respectively.  
	
	These equations can also be regarded as a mixture of TBA and NLIE, but these equations have the 
	advantage that the number of unknown functions is less than it is in the TBA equations, hence these equations are much
	more convenient for numerical studies. Another advantage is that if the model has a single coupling constant then
	the equations  are valid for all real value of the coupling constant.
	
	Later two component nonlinear integral equations
	were derived for the ground state energy of the $O(4)$ nonlinear $\sigma$-model \cite{HA}
	in the context of the light-cone approach \cite{LC,LCS}, formulating the $O(4)$ nonlinear $\sigma$-model as the
	$S \to \infty$ limit of the inhomogeneous spin-S vertex model,
	and
	by modifying the kernel functions of these equations similar nonlinear integral equations
	were proposed for the finite size effects in the SS-model for finite values of the couplings \cite{HA}.
	 
	In this paper, formulating $\cal{M}$$A^{+}_{k,r}$ model as the appropriate continuum limit of the critical,
	 inhomogeneous
	 RSOS($k,q$) lattice model, we derive two component nonlinear integral equations for describing the ground 
	 state energy of the model. The lattice version of these equations describe the finite size effects in the
	 corresponding critical RSOS($k,q$) lattice model. 
	 
	 When the allowed values of the heights tends to infinity ($r \to \infty$), we obtain the equations
         corresponding to the current-current perturbation of the $SU(2)_k$ WZW model. The lattice version 
	 of these equations describe the finite size effects in the isotropic spin-$k/2$ vertex model.
	
	Then formally releasing the restriction in the two component nonlinear integral equations of the
	$\cal{M}$$A^{+}_{k,r}$ model, we propose
	two component equations for the FSSG models, the lattice version of which correspond to the case of the 
	anisotropic higher spin vertex model. 
	
	Due to the minimal number of components, our new nonlinear integral equations are more convenient for 
	numerical studies than the corresponding TBA equations. Moreover, the minimality of the number of components makes
	the treatment of the excited state problem much easier than in the context of the TBA approach.


    The paper is organized as follows. In section 2, we recall the main results of the light-cone lattice approach 
    to the $\cal{M}$$A^{+}_{k,r}$ models.
	In section 3, we summarize the most important properties of the critical RSOS($k,q$) lattice models, comprising
	the T- and Y-systems, which follow from the Bethe Ansatz solution of the model. 
	In section 4, using the light-cone lattice approach and  Suzuki's \cite{suz} method
	 we derive nonlinear integral equations for describing the finite size effects. 
    In section 5, formulating the RSOS($k,q$) lattice model as an RSOS($k',q'$) model with
    $k'=r-k-2$, $q'=r-q-2$, we derive another nonlinear 
	integral equations for the same model.
    In section 6, we glue together the two nonlinear integral equations to a much 
	simpler equation. This equation describes the finite-volume ground state energy of our perturbed coset model.
	In section 7, we perform some analytical and numerical tests on our two component equations.
	In section 8, we propose nonlinear integral equations for describing the finite-volume ground state energy of 
	the fractional supersymmetric sine-Gordon models by formally releasing the restriction in the equations
	obtained for the $\cal{M}$$A^{+}_{k,r}$ models.
	In section 9, we briefly summarize how the two component nonlinear integral equations look like in the
	case of the solvable lattice models.
	The summary and conclusions of this paper are given in section 10.


\section{The light-cone lattice approach to the $\cal{M}$$A^{+}_{k,r}$ models}

In the following we use the integrable light-cone lattice regularization of our model \cite{SR,LC}. 
 Starting from the inhomogeneous critical RSOS lattice model, in \cite{SR} a local integrable 
 lattice regularization of the Hamiltonian of the $\cal{M}$$A^{+}_{k,r}$ model was proposed.
 As a result of the locality of the lattice Hamiltonian, 
 in the continuum limit of this regularization, one obtains the $\cal{M}$$A^{+}_{k,r}$ model, plus
 a decoupled $SU(2)_k \times SU(2)_{r-k-2}/SU(2)_{r-2}$ coset conformal field theory.
 In the following to avoid such \lq\lq particle doublings", following the lines of \cite{LC},
 we define a nonlocal Hamiltonian for the $\cal{M}$$A^{+}_{k,r}$ model from the same lattice model.
 The final result of the two regularizations are the same in the continuum.
 The fields of the regularized theory
are defined at sites (\lq\lq events") of a light-cone lattice and the
dynamics of the system is defined by 
translations in the left and right light-cone directions. These are given by inhomogeneous
transfer matrices of the RSOS($k,q$) model considered on the critical line
separating the regimes III and IV.

This approach is particularly useful for
calculating the finite size dependence of physical quantities.
We take $N$ points ($N$ is even) in the spatial direction and use periodic
boundary conditions. The lattice spacing is related to $l$, the
(dimensionful) size of the system :
\begin{equation}
a=\frac{2l}{N}.
\end{equation}


  The energy ($E$) and momentum ($P$) of the physical states can be
obtained from the eigenvalues of the light-cone transfer matrices (\ref{14}):
\begin{equation}
e^{i \frac{a}{2}(H+P)}=\frac{T^{(k)}_k (x_0+i)}{T^{(k)}_0 (x_0+i(k+1))}, 
\qquad \qquad 
e^{i \frac{a}{2}(H-P)}=\frac{T^{(k)}_k (-x_0-i)}{T^{(k)}_0 (-x_0-i(k+1))}. \label{6}
\end{equation}
Besides the usual procedure, taking the thermodynamic limit
($N \to \infty$) first, followed by the continuum limit ($a \to 0$)
one can also study continuum limit in finite volume by taking
$N \to \infty$ and tuning the inhomogeneity parameter $x_0$
simultaneously as
\begin{equation}
x_0=\frac{2}{\pi}\log \frac{4}{{m}a}=\frac{2}{\pi}\log \frac{2N}{ml},
\label{7}
\end{equation}
where the mass parameter ${m}$ is the infinite volume mass gap of the theory. 
If we take this continuum limit we get the energy and momentum eigenvalues of the
$\cal{M}$$A^{+}_{k,r}$ models.

%


\section{RSOS($k,q$) lattice models}

The restricted solid-on-solid models of Andrews, Baxter and Forrester \cite{ABF} are IRF or interaction-round-a-face models
 \cite{Bax}.
The heights or spins $a,b,c,d,$ etc. at each site of the square lattice take the values 1,2,...,r-1.
These spins are subject to the nearest neighbour constraint $a-b=\pm 1$ for each pair of adjacent spins $a,b$.
The spins therefore take values on the Dynkin diagram of the classical Lie algebra $A_{r-1}$. The statistical weight
assigned to an elementary face of the lattice 
is zero unless the spins around the face satisfy the adjacency rules described later.
The weights of allowed faces are given by
\begin{equation}
W \left( \begin{array}{cc|} d & c \\ a & b \end{array} \ u \right) 
=\frac{\sin(\lambda-u)}{\sin(\lambda)} \ \delta_{ac}+\frac{\sin(u)}{\sin(\lambda)} \
\sqrt{\frac{S_a S_c}{S_b S_d}} \ (-1)^{(a-c)/2} \ \delta_{bd},
\end{equation}
where
\begin{equation}
\lambda=\frac{\pi}{r}, \qquad S_a=\sin(a\lambda).
\end{equation}
The spectral parameter $u$ is related to the spatial anisotropy, $\lambda$ is the crossing parameter and
$\delta$ is the Kronecker delta. 

The critical RSOS($k,q$) lattice models are obtained by fusing $k \times q$ blocks of face weights together. 
In the following we consider the model on the regime III/IV critical line.
The face weights of the fused RSOS($k,q$) models, with max$(k,q)\leq r$, are obtained by the fusion process \cite{JMO}.
The RSOS($k,q$) face weights are given explicitly by \cite{JMO,BZR}
\begin{eqnarray}
W^{k,q} \left( \begin{array}{cc|} a_{q+1} & b_{q+1} \\ a_1 & b_1 \end{array} \ u \right)= 
\prod_{n=0}^{q-2} \prod_{j=0}^{q-1} \frac{\sin(\lambda)}{\sin(u+(n-j)\lambda)} \nonumber \\
\sum_{a_1,...,a_q} \prod_{n=1}^{q} 
W^{k,1} \left( \begin{array}{cc|} a_{n+1} & b_{n+1} \\ a_n & b_n \end{array} \ u+(k-1)\lambda \right)
\end{eqnarray}
independent of the values of the edge spins $b_2,..,b_q$, where the $k \times 1$ face weights are given in turn by
\begin{equation}
W^{k,1} \left( \begin{array}{cc|} b_{1} & b_{k+1} \\ a_1 & a_{k+1} \end{array} \ u \right)=
\sum_{a_2,...,a_k} \prod_{n=1}^{k} 
W\left( \begin{array}{cc|} b_{n} & b_{n+1} \\ a_n & a_{n+1} \end{array} \ u+(n-k)\lambda \right)
\end{equation}
independent of the values of the edge spins $b_2,..,b_k$. For the RSOS($k,q$) models, adjacent spins or heights are
subject to constraints
\begin{equation}
0\leq(a_i-a_j+m)/2 \leq m, \qquad m<a_i+a_j<2r-m,
\end{equation}
where $m$ is equal to $k$ for a horizontal pair and $q$ for a vertical pair.

The RSOS($k,q$) lattice models are exactly solvable. In particular, the fused face weights satisfy
the generalized Yang-Baxter equation
\begin{eqnarray}
\sum_{g} W^{k,q} \left( \begin{array}{cc|} f & g \\ a & b \end{array} \ u \right)
W^{k,s} \left( \begin{array}{cc|} e & d \\ f & g \end{array} \ u+v \right)
W^{q,s} \left( \begin{array}{cc|} d & c \\ g & b \end{array} \ v \right) \nonumber \\
=\sum_{g} W^{q,s} \left( \begin{array}{cc|} e & g \\ f & a \end{array} \ v \right)
W^{k,s} \left( \begin{array}{cc|} g & c \\ a & b \end{array} \ u+v \right)
W^{k,q} \left( \begin{array}{cc|} e & d \\ g & c \end{array} \ u \right)
\end{eqnarray}
This is an immediate consequence of the elementary Yang-Baxter equation satisfied by the $1 \times 1$
face weights and leads to commuting transfer matrices. Suppose that ${\bf a}$ and ${\bf b}$ are
allowed spin configurations of two consecutive rows of an $N$ column lattice with periodic boundary
conditions. Assuming that $N$ is even the matrix elements of the RSOS($k,q$) transfer matrix 
with alternating inhomogeneities are given by
\begin{equation}
\langle {\bf a}| T^{(k)}_q(x)|{\bf b} \rangle=
\prod_{j=1}^{N} 
W^{k,q} \left( \begin{array}{cc|} b_j & b_{j+1} \\ a_j & a_{j+1} \end{array} \ 
\frac{i \pi (x-(-1)^{j+1}x_0)}{2 r}
+\frac{k+1-q}{2} \lambda \right) \label{14}
\end{equation}
where $a_{N+1}=a_1$ and $b_{N+1}=b_1$, and $x_0$ is the inhomogeneity, which is zero if one is interested in the
finite size dependence of the energy levels of the lattice model, and 
must be tuned according to (\ref{7})
to obtain a continuum field theory 
in finite volume.
For a fixed value of $k$ the Yang-Baxter equations imply the commutation relations
\begin{equation}
T^{(k)}_q(x) \ T^{(k)}_{q'}(y)=T^{(k)}_{q'}(y) \ T^{(k)}_{q}(x).
\end{equation}
The eigenvalues of these mutually commuting transfer matrices satisfy the so-called
T-system functional relations \cite{KPA}
\begin{equation}
T^{(k)}_{p}(x-i) \ T^{(k)}_{p}(x+i)=f^{(k)}_{p}(x)+T^{(k)}_{p-1}(x) \ T^{(k)}_{p+1}(x), \label{16}
\end{equation}
where
\begin{equation}
f_{p}^{(k)}=T_0^{(k)}(x+i(p+1)) \ T_0^{(k)}(x-i(p+1)),
\end{equation}
and
\begin{equation}
T_{-1}^{(k)}=0, \qquad \qquad T_0^{(k)}(x)=\prod_{j=0}^{k-1} \phi(x+i(k-1-2j)), 
\end{equation}
\begin{equation}
\phi(x)=\frac{\sinh^{N/2}(\frac{\lambda}{2}(x-x_0)) \ \sinh^{N/2}(\frac{\lambda}{2}(x-x_0))}{\sin^N(\lambda)}.
\end{equation}
These T-system equations are the same as the ones of the anisotropic higher-spin vertex model \cite{BZR}.
The RSOS($k,q$) transfer matrices satisfy the symmetry \cite{BZR}
\begin{equation}
T^{(k)}_{q}(x)={\bf Y} T^{(k)}_{r-2-q}(x-ir), \qquad q=1,2,...,r-3, \label{20}
\end{equation}
where ${\bf Y}$ is the height reflection operator
\begin{equation}
\langle {\bf a}|{\bf Y}|{\bf b} \rangle=\prod_{j=1}^{N} \delta_{a_j}^{r-b_j},
\qquad [T^{(k)}_{q}(x),{\bf Y}]=0.
\end{equation}
It follows that the fusion hierarchy closes with $T^{(k)}_{r-1}(x)=0$.
The functions $T^{(k)}_{q}(x)$ are all periodic on the complex plane with period $2ir$
\begin{equation}
 T^{(k)}_{q}(x)=T^{(k)}_{q}(x+2ir).
\end{equation}
It will be important for our later considerations
 that it can be shown \cite{BZR}
  that the transfer matrices $T^{(k)}_{q}(x)$ (\ref{14}) of an RSOS($k,q$) model differs from the transfer matrices $T^{(k')}_{q'}(x)$
   of an RSOS($k',q'$) model only in some trivial normalizations, if $k'=r-k-2$, and $q'=r-q-2$:
\begin{equation}
T^{(k)}_{q}(x)\sim T^{(k')}_{q'}(x), \qquad k'=r-k-2, \quad q'=r-2-q. \label{23}
\end{equation}

The solution of the T-system (\ref{16}) can be characterized by a set of Bethe roots \{$x_j, j=1,...,kN/2$ \}, which
are the solutions of the Bethe Ansatz equations \cite{BZR}
\begin{equation}
\frac{T_0^{(k)}(x_j+i)}{T^{(k)}_0(x_j-i)}=-e^{2i\omega} \frac{Q^{(k)}(x_j+2i)}{Q^{(k)}(x_j-2i)}, 
\qquad j=1,...,kN/2. \label{24}
\end{equation}
where
\begin{equation}
Q^{(k)}(x)=\prod_{j=1}^{kN/2} \sinh\left(\frac{\lambda}{2}(x-x_j)\right), \label{25}
\end{equation}
\begin{equation}
e^{i\omega r}=-(-1)^{kN/2} Y, \qquad \omega\neq 0. \label{26}
\end{equation}
where $Y=\pm 1$ is the eigenvalue of the operator ${\bf Y}$. 
The eigenvalues of the transfer matrices (\ref{14}) are of the form \cite{BZR,KPA}
\begin{equation}
T^{(k)}_p(x)=\sum_{l=1}^{p+1} {\lambda}^{(p)}_{l} (x), 
\end{equation}
where
\begin{equation}
{\lambda}^{(p)}_{l}(x)= T^{(k)}_0(x+i(2l-p-2)) \ \frac{e^{i \omega (p+2-2l)} \ Q^{(k)}(x+i(p+1)) \ Q^{(k)}(x-i(p+1)) }
{Q^{(k)}(x+i(2l-p-1)) \ Q^{(k)}(x+i(2l-p-3)) }. 
\end{equation}
From a T-system (\ref{16}) one can define a Y-system as follows
\begin{equation}
y^{(k)}_j(x)=\frac{T^{(k)}_{j-1}(x) T^{(k)}_{j+1}(x) }{f^{(k)}_j(x)}, \label{29}
\end{equation}
\begin{equation}
Y^{(k)}_j(x)=1+y^{(k)}_j(x)=\frac{ T^{(k)}_j (x+i) T^{(k)}_j (x-i) }{ f^{(k)}_j(x) }. \label{30}
\end{equation}
These functions satisfy the Y-system equations \cite{ZamiY,Kuniba,KP}
\begin{equation}
y^{(k)}_j(x+i) y^{(k)}_j(x-i)=Y^{(k)}_{j-1}(x) Y^{(k)}_{j+1}(x), \qquad j=1,...r-3. \label{31}
\end{equation}
It follows from (\ref{20}) that the Y-system (\ref{29}-\ref{31}) also closes (ie. $y^{(k)}_{r-2}(x)=0$).
One can redefine the T-system elements with
\begin{equation}
T^{(k)}_j(x) \rightarrow \tilde{T}^{(k)}_j(x)=\sigma_j(x) T^{(k)}_j(x),
\end{equation}
where $\sigma_j(x)$ satisfies the relation
\begin{equation}
\sigma_j(x+i) \sigma_j(x-i)=\sigma_{j-1}(x) \sigma_{j+1}(x),
 \end{equation}
then the new $\tilde{T}^{(k)}_j(x)$ functions satisfy the same T-system relations as (\ref{16})
 but with different $f^{(k)}_j(x)$ functions: 
\begin{equation}
f^{(k)}_j(x) \rightarrow \tilde{f}^{(k)}_j(x)=\sigma_j(x+i) \sigma_j(x-i) f^{(k)}_j(x).
\end{equation}
This is called gauge transformation. Under such a transformation the Y-system (\ref{29}-\ref{31}) is invariant.
As we will see later the energy of  the model can be expressed by an element of the gauge invariant Y-system
(see eq. (\ref{gaugeY}) later).
On the other hand one has the freedom of choosing that gauge for the T-system (\ref{16}) which is convenient for 
the particular calculations.
 For later use it is important to note that the Y-system (\ref{29}-\ref{31}) of an 
 RSOS($k,q$) model following from (\ref{23}) is identical to the one of an RSOS($k',q'$) model ie.
 \begin{equation}
 y^{(k)}_q (x)=y^{(k')}_{q'} (x), \qquad k'=r-k-2 \qquad q'=r-q-2. \label{35}
 \end{equation}
From this point of view an RSOS($k,q$) model is equivalent to an RSOS($k',q'$) model, and 
so in any particular calculation we can choose the formulation which is more convenient.




\section{Nonlinear integral equations I.}

In this section we apply Suzuki's method \cite{suz} to obtain a set of nonlinear integral equations
 for the ground state energy of our model in finite
volume. Consider the following gauge transformation of the T-system (\ref{16})
\begin{equation} 
\tilde{T}^{(k)}_p(x+i)\tilde{T}^{(k)}_p(x-i)=\tilde{f}^{(k)}_p (x)+\tilde{T}^{(k)}_{p-1}(x)\tilde{T}^{(k)}_{p+1}(x), \label{36}
\end{equation}
where
\begin{equation}
\tilde{f}^{(k)}_p(x)=\prod_{j=1}^{p} \phi(x+i(p-k-2j)) \ \phi(x-i(p-k-2j)),
\end{equation}  
and
\begin{equation}
\tilde{T}^{(k)}_{-1}(x)=0, \qquad \qquad \tilde{T}^{(k)}_0(x)=1.
\end{equation}
The solutions of these equations are  of the form \cite{suz} 
\begin{equation}
\tilde{T}^{(k)}_p(x)=\sum_{l=1}^{p+1} \tilde{\lambda}_{l}^{(p)}(x),
\end{equation}
where
\begin{equation}
\tilde{\lambda}_{l}^{(p)}(x)=e^{i \omega (p+2-2l)} \
\tilde{\psi}_{l}^{(p)}(x)\ \frac{Q^{(k)}(x+i(p+1)) \ Q^{(k)}(x-i(p+1)) }{Q^{(k)}(x+i(2l-p-1)) \ Q^{(k)}(x+i(2l-p-3)) },
\end{equation}
\begin{equation}
\tilde{\psi}_{l}^{(p)}(x) = \prod_{j=1}^{p-l+1} \phi(x+i(p-k-2j+1)) \prod_{j=1}^{l-1} \phi(x-i(p-k-2j+1)). \label{41}
\end{equation}
We define the following auxiliary functions \cite{suz}
\begin{equation}
y^{(k)}_j(x)=\frac{\tilde{T}^{(k)}_{j-1}(x) \tilde{T}^{(k)}_{j+1}(x) }{\tilde{f}^{(k)}_j(x)}, \ \qquad \qquad j=1,\dots,k
\label{42}
\end{equation}
\begin{equation}
Y^{(k)}_j(x)=1+y^{(k)}_j(x)=\frac{ \tilde{T}^{(k)}_j (x+i) \tilde{T}^{(k)}_j (x-i) }{\tilde{ f}^{(k)}_j(x) }, \ \qquad \qquad j=1,\dots,k
 \label{43}
\end{equation}
\begin{equation}
b(x)=\frac{ \tilde{\lambda}_{1}^{(k)}(x+i)+\dots+\tilde{\lambda}_{k}^{(k)}(x+i) }{\tilde{\lambda}_{k+1}^{(k)}(x+i) },
\qquad \qquad \mathcal{B}(x)=1+b(x), \label{44}
\end{equation}
\begin{equation}
\bar{b}(x)=\frac{ \tilde{\lambda}_{2}^{(k)}(x-i)+\dots+\tilde{\lambda}_{k+1}^{(k)}(x-i) }{\tilde{\lambda}_{1}^{(k)}(x-i) },
\qquad \qquad \bar{\mathcal{B}}(x)=1+\bar{b}(x). \label{45}
\end{equation}
From (\ref{36}-\ref{41}) it follows that the auxiliary functions (\ref{42}-\ref{45}) satisfy the following functional relations:
\begin{equation}
\tilde{T}^{(k)}_k(x+i)=e^{-ik \omega} \ \prod_{j=1}^{k} \phi(x+2ij) \ \frac{Q^{(k)}(x-ik) }{ Q^{(k)}(x+ik) } \ \mathcal{B}(x),
 \label{46}
\end{equation}
\begin{equation}
\tilde{T}^{(k)}_k(x-i)=e^{ik \omega} \ \prod_{j=1}^{k} \phi(x-2ij) \ \frac{Q^{(k)}(x+ik) }{ Q^{(k)}(x-ik) } \ \bar{\mathcal{B}}(x),
 \label{47}
\end{equation}
\begin{equation}
b(x)=e^{i(k+1) \omega} \ \frac{ \phi(x) }{ \prod_{j=1}^{k} \phi(x+2ij) } \  \frac{ Q^{(k)}(x+ik+2i) }{ Q^{(k)}(x-ik) } \ \tilde{T}^{(k)}_{k-1}(x), \label{}
\end{equation}
\begin{equation}
\bar{b}(x)=e^{-i(k+1) \omega} \ \frac{ \phi(x) }{ \prod_{j=1}^{k} \phi(x-2ij) } \  \frac{ Q^{(k)}(x-ik-2i) }{ Q^{(k)}(x+ik) } \ \tilde{T}^{(k)}_{k-1}(x), \label{}
\end{equation}
\begin{equation}
\mathcal{B}(x) \bar{\mathcal{B} }(x)=Y^{(k)}_k(x), \label{50}
\end{equation}
\begin{equation}
y^{(k)}_j(x+i) y^{(k)}_j(x-i)=Y^{(k)}_{j-1}(x) Y^{(k)}_{j+1}(x) \qquad \qquad j=1,\dots,k-2, \label{51}
\end{equation}
\begin{equation}
y^{(k)}_{k-1}(x+i) y^{(k)}_{k-1}(x+i) = Y^{(k)}_{k-2}(x) \mathcal{B}(x) \bar{\mathcal{B} }(x),
\end{equation}
\begin{equation}
\tilde{T}^{(k)}_{k-1}(x+i) \tilde{T}^{(k)}_{k-1}(x-i)=\tilde{f}^{(k)}_{k-1}(x) Y^{(k)}_{k-1}(x). \label{53}
\end{equation}
We introduce two other auxiliary functions
\begin{equation}
\Psi^{(k)}_1(x)=Q^{(k)}(x-ik), \qquad \qquad \Psi^{(k)}_2(x)=Q^{(k)}(x+ik)=\Psi^{(k)}_1(x-2i(r-k)). \label{54}
\end{equation}
In order to be able to derive integral equations from these functional relations one needs to know the positions 
of the zeroes and the poles of the auxiliary functions (\ref{42}-\ref{45}). Due to the relations (\ref{46}-\ref{53}) we only 
need to know the zeroes of $Q^{(k)}(x)$ and $\tilde{T}^{(k)}_j(x)$. For the ground state of our model the zeroes of  $Q^{(k)}(x)$ form
 $\frac{N}{2}$ pieces  of $k$-strings, but for  finite $N$ values there are deviations from the string hypothesis \cite{str}. 
From \cite{str} one can see that Bethe roots having the largest deviations from the imaginary positions prescribed
by the string hypothesis are those that would have imaginary parts $\pm (k-1)$ according to the string hypothesis. 
These deviations are always less
than $1/2$ and this is important 
 because the analytic properties of $\Psi^{(k)}_1(x)$ ($\Psi^{(k)}_2(x)$) are influenced mainly by these
roots on the upper (lower) half plane of the complex plane near the real axis. 

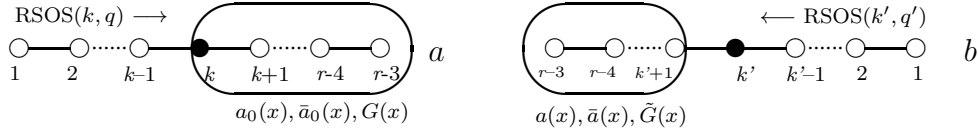
\begin{figure}[htbp]
\begin{center}
\begin{picture}(280,30)(-140,-15)
\put(-163,-7) {\usebox{\RSNI}}
\put(-74,-7) {\usebox{\RSNII}}
\put(-130,-15){\parbox{130mm}{\caption{ \label{2f}\protect {\small
Graphical representation of nonlinear integral equations $a$.) I and $b$.) II. }}}}
\end{picture}
\end{center}
\end{figure}

  The transfer matrices $\tilde{T}^{(k)}_j(x) \quad j=1,\dots,k$ 
(\ref{36}) have no zeroes in the ground state in the \lq\lq main" strip $0\leq|\mbox{Im}x | \leq 1$.  We can list the strips
where the auxiliary functions are analytic and non zero (ANZ).
\begin{eqnarray}
\Psi^{(k)}_1(x) \qquad  &\mbox{ANZ}& \qquad -2(r-k)-1/2 \leq \mbox{Im}\ x \leq 1/2, \ \ \mbox{mod}  \ 2r  \nonumber \\
\Psi^{(k)}_2(x) \qquad &\mbox{ANZ}& \qquad  -1/2 \leq \mbox{Im}\ x \leq 2(r-k)+1/2, \ \ \mbox{mod}  \ 2r \nonumber  \\
b(x),\mathcal{B}(x) \qquad &\mbox{ANZ}& \qquad 0 < |\mbox{Im}\ x| \leq 1/2, \ \ \mbox{mod}  \ 2r  \nonumber  \\ 
\bar{b}(x),\bar{\mathcal{B}}(x) \qquad &\mbox{ANZ}& \qquad 0 < |\mbox{Im}\ x| \leq 1/2, \ \ \mbox{mod}  \ 2r \label{55}\\
y^{(k)}_j(x)  \qquad &\mbox{ANZ}& \qquad 0 \leq |\mbox{Im}\ x| \leq 1, \ \ \mbox{mod}  \ 2r\qquad j=1,\dots,k-1,   \nonumber  \\
Y^{(k)}_j(x) \qquad &\mbox{ANZ}& \qquad 0 \leq |\mbox{Im}\ x| \leq \epsilon, \ \ \mbox{mod}  \ 2r \qquad j=1,\dots,k-1 \quad \epsilon>0,   \nonumber  \\
\tilde{T}^{(k)}_j(x)  \qquad &\mbox{ANZ}& \qquad 0 \leq |\mbox{Im}\ x| \leq 1, \ \ \mbox{mod}  \ 2r \qquad j=1,\dots,k. \nonumber 
\end{eqnarray}
We introduce new variables by shifting the arguments of $b(x),\mathcal{B}(x)$ and $\bar{b}(x),\bar{\mathcal{B}}(x)$ by $\pm i \gamma$ \cite{suz} 
\begin{equation}
a_0(x)=b(x-i\gamma), \qquad \qquad U_0(x)=\mathcal{B}(x-i\gamma)=1+a_0(x),
\end{equation}
 \begin{equation}
\bar{a}_0(x)=\bar{b}(x+i\gamma), \qquad \qquad 
\bar{U}_0(x)=\bar{\mathcal{B}}(x+i\gamma)=1+\bar{a}_0(x),
\end{equation}
where $0<\gamma<1/2$ is an arbitrary real parameter. This shift is necessary because the original functions
$b(x),\bar{b}(x)$ have zeroes on the real axis.
Due to the ANZ property of $\tilde{T}^{(k)}_k(x)$ (\ref{55}) and the fact that $\lim_{|x| \to \infty} \frac{d^2}{dx^2} \log  
\tilde{T}^{(k)}_k(x) = 0$ the following relation holds \cite{suz}
\begin{equation}
0=\int\limits_{-\infty}^{\infty}dx \ \frac{d^2}{dx^2} \log  \tilde{T}^{(k)}_k(x-i)\ e^{iq(x-i)}-\int\limits_{-\infty}^{\infty} dx \
\frac{d^2}{dx^2} \log  \tilde{T}^{(k)}_k(x+i)\ e^{iq(x+i)}. \label{58}
\end{equation}
From (\ref{46}) and (\ref{47}) one can express $\tilde{T}^{(k)}_k(x \pm i)$ with the auxiliary functions and by substituting these 
expressions  into (\ref{58}) one gets  in  Fourier space 
\begin{eqnarray}
\tilde{d^2 l} \Psi^{(k)}_1 (q) &=& \pi N q e^{(r-k)q} \ \frac{ \sinh (kq) \ \cos (x_0 q) }{ \sinh (2q) \ \sinh (rq) } \nonumber \\
                                      &+& \frac{ e^{(r-k+1-\gamma)q } \ \ \tilde{d^2 l} \bar{U}_0 (q)}{4 \ \cosh (q) \ \sinh[(r-k)q] } - 
                                                  \frac{ e^{(r-k-1+\gamma)q } \ \ \tilde{d^2 l} {U}_0 (q) }{4 \ \cosh (q) \ \sinh[(r-k)q] },  \\
\tilde{d^2 l} \Psi^{(k)}_2 (q) &=& \pi N q e^{-(r-k)q} \ \frac{ \sinh (kq) \ \cos (x_0 q) }{ \sinh (2q) \ \sinh (rq) } \nonumber \\
                                      &+& \frac{ e^{-(r-k+1+\gamma)q } \ \tilde{d^2 l} \bar{U}_0 (q) }{4 \ \cosh (q) \ \sinh[(r-k)q] }  - 
                                                  \frac{ e^{-(r-k-1-\gamma)q } \ \ \tilde{d^2 l} {U}_0 (q) }{4 \ \cosh (q) \ \sinh[(r-k)q] },  
\end{eqnarray}
where we introduced the notation
\begin{equation}
\tilde{d^2 l} F(q)=\int\limits_{-\infty}^{\infty} dx \ e^{iqx} \ \frac{d^2}{d^2 x} \log F(x).
\end{equation}
One can derive similar relations for $\tilde{d^2 l} y^{(k)}_j(q)$'s and  $\tilde{d^2l} Y^{(k)}_j(q)$'s from (\ref{51}), 
and $\tilde{d^2 l} \tilde{T}^{(k)}_{k-1}(q)$
and $\tilde{d^2 l} Y^{(k)}_{k-1}(q)$ from (\ref{53}). Substituting these relations into the definitions of $a_0(x)$ and $\bar{a}_0(x)$, one obtains $k+1$ algebraic relations in Fourier space. After taking the inverse Fourier
transformation of these relations and integrating twice over $x$ determining the integration constants from the $x \to \infty$
asymptotics of the auxiliary functions and exploiting that following from (\ref{26}) $\omega=\pi/r$ for the ground state,
 we get Suzuki's \lq\lq half TBA-NLIE" type of equations \cite{suz}
\begin{eqnarray}
\log y^{(k)}_1(x) &=& (K*\log Y^{(k)}_2)(x), \nonumber \\  
\log y^{(k)}_j (x) &=& (K* \log Y^{(k)}_{j-1})+(K*\log Y^{(k)}_{j+1})(x), \qquad  \quad j=2,\dots,k-2,  \nonumber  \\   
\log y^{(k)}_{k-1}(x) &=&  (K*\log Y^{(k)}_{k-2})(x)+ (K^{+\gamma}*\log U_0)(x)+
                                         (K^{-\gamma}*\log \bar{U}_0)(x),    \nonumber  \\
\log a_0(x) &=& \mathcal{D}_N (x-i \gamma)+ (G*\log  U_0)(x)- (G^{+2(1-\gamma)}*\log \bar{U}_0 )(x)  \label{62} \\
                            &+& (K^{-\gamma}*\log Y^{(k)}_{k-1})(x)+i \frac{\pi}{r-k}-i \delta \pi,    \nonumber  \\
\log \bar{a}_0(x) &=& \mathcal{D}_N (x+i \gamma)+ (G*\log  \bar{U}_0)(x)- (G^{-2(1-\gamma)}*\log  U_0 )(x)    \nonumber  \\
                             &+& (K^{+\gamma}*\log Y^{(k)}_{k-1})(x)-i \frac{\pi}{r-k}+i \delta \pi,   \nonumber
\end{eqnarray}
where $(K*f)(x)=\int dy \ K(x-y)\ f(y) $ is the convolution and the \lq\lq source" function $\mathcal{D}_N (x)$ on the lattice 
reads as:
\begin{equation}
\mathcal{D}_N (x)= i N \arctan \left[    \frac{ \sinh \left(  \frac{\pi (x+i)}{2}   \right)  }{ \cosh \left(\frac{\pi x_0}{2} \right)} \right] ,
\qquad \qquad x_0=\frac{2}{\pi} \log \left(    \frac{2N}{ml}  \right),
\end{equation}
and
\begin{equation}
\delta=(N/2)_{mod \ 2} \in \{0,1\}
\end{equation}
 the value of which is zero ($\delta \equiv 0$) for the ground state of the 
$\cal{M}$$A^{+}_{k,r}$ model.
The kernel functions of (\ref{62}) are of the form 
\begin{equation}
K(x)=\frac{1}{4 \ \cosh (\pi x /2) }, \label{64}
\end{equation}
\begin{equation}
G(x)=\int\limits_{-\infty}^{\infty} \frac{dq}{2 \pi} \ e^{iqx} \ \frac{\sinh[(r-k-1)q]}{2 \ \cosh(q) \ \sinh[(r-k)q] },
\end{equation}
and we have used the notation
\begin{equation}
f^{\pm \eta}(x)=f(x \pm i \eta).
\end{equation}
We get the continuum limit of eqs. (\ref{62}) by taking the $N \to \infty$ limit. The kernel functions
do not change because they are independent of $N$, but the \lq\lq source" function changes  in the continuum as
\begin{equation}
\mathcal{D}(x)=\lim_{N \to \infty} \mathcal{D}_N (x)=-ml \cosh \left( \frac{\pi x}{2} \right). \label{67}
\end{equation}
In the TBA language the complex auxiliary functions $a_0(x)$ and $\bar{a}_0(x)$ resum the contributions of 
those Y-system elements whose index is larger than $k-1$. Equations (\ref{62}) are graphically represented in figure \ref{2f}$a$.

 The big \lq\lq bubble" denotes the complex auxiliary functions which resum the contributions of those
TBA nodes, which are inside it. In our notation the names of the complex unknown
functions and the kernel functions are indicated. It is interesting to recognize that in (\ref{62}) in the equations
of the complex unknown functions,
the only parameter which appears in $G(x)$ and in the integration constants, is $r-k$, which is nothing but
the \lq\lq length" of that part of the TBA diagram the contribution of which these complex functions resum. 



The energy and momentum of the model (\ref{6}) can be easily expressed by our auxiliary functions due to the fact
that  apart from some trivial normalization factors: 
\begin{equation}
e^{i \frac{a}{2}(H+P)} \sim \tilde{T}^{(k)}_k(x_0+i),  \qquad \mbox{and} \qquad e^{i \frac{a}{2}(H-P)} \sim \tilde{T}^{(k)}_k(-x_0-i).
\nonumber
\end{equation}
After some straightforward calculations \cite{Kuniba,SGg} one gets 
the following result for the ground state energy in the continuum limit
\begin{eqnarray}
E_0(l) &=& E_{bulk}-\frac{m}{4} \int\limits_{-\infty}^{\infty} dx  \ \cosh \left( \frac{\pi (x-i \gamma)}{2}\right) \log U_0(x)
\label{68} \\
&-& \frac{m}{4} \int\limits_{-\infty}^{\infty} dx \
\cosh \left( \frac{\pi (x+i \gamma)}{2}\right) \log \bar{U}_0(x) , \nonumber
 \end{eqnarray}
where $E_{bulk}$ is an overall divergent factor in the continuum and it has the form on the lattice:
\begin{equation}
E_{bulk}=\frac{N^2}{2l} \ \sum_{j=0}^{k/2-1} \ \frac{1}{i} \ \log\left[  
\frac{\sinh\left[ \lambda(i(1+2j)-x_0) \right]}{ \sinh\left[\lambda(i(1+2j)+x_0) \right]  }  \right]
\qquad  \quad \mbox{when}\  k \ \mbox{is even,}
\end{equation}
\begin{equation}
E_{bulk}=\frac{N^2}{2l} \left(  \chi(2x_0) +  
\sum_{j=0}^{(k-3)/2} \ \frac{1}{i}  \log\left[  
\frac{\sinh\left[ \lambda(2i(1+j)-x_0 \right]}{ \sinh\left[ \lambda(2i(1+j)+x_0 \right]}  \right]  \right)
\quad  \mbox{when}\ k \ \mbox{is odd,}
\end{equation}
where
\begin{equation}
\chi(x)=  \int\limits_{-\infty}^{\infty} \frac{dq}{2 \pi} \  \frac{\sin(qx)}{q}
\frac{\sinh[(r-1)q]}{2 \ \cosh(q) \ \sinh(rq) }.
\end{equation}
From these formulas one can easily see that $E_{bulk}$ is independent of the Bethe roots, thus independent of the dynamics
of the model and so $E_{bulk}$ appears in the energy expression as a bulk energy term same for all states of the model. 
One can easily see that $E_{bulk}$ is divergent in the continuum, but after subtracting this divergent term from the energy  
the finite integral terms in (\ref{68}) describe the finite size dependence of the ground state energy of the model.
We note that eqs. (\ref{62}) with the energy expression (\ref{68}) are nothing else but the equations proposed by Dunning \cite{dun}
for the $\cal{M}$$A^{+}_{k,r}$ models. However, here these equations are derived
 from the appropriate lattice regularization of these models.
  
  At this point we have to make an important remark, about the value of $\gamma$. According to Dunning's conjecture
the value of the parameter $\gamma$ can tend to 1, just as in the case of the sine-Gordon NLIE.
Nevertheless from our derivation one can see, that because of the nonzero deviations of the $k$-strings, $\gamma$ 
cannot be larger than $1-\epsilon$, where $\epsilon$ is the largest deviation of the $k$-strings from the line 
$\mbox{Im}(x)=k-1$, and because it is well known that $\epsilon$ is always finite, and not infinitesimal \cite{str}, 
$\gamma$ cannot tend to 1, and it must be kept finite. This statement can be checked numerically as well.
Namely if one lowers the value of $\gamma$ in the numerical calculations, below a critical value of this parameter
the numerical value of the ground state energy changes.
 In our equations we assumed
that these string deviations are always less than $1/2$ at any values of $N$ (or $l$ in the continuum), 
that is why we restricted the allowed values of $\gamma$ into the $(0,1/2)$ interval.




\section{Nonlinear integral equations II.}

In this section we derive a new set of nonlinear integral equations to describe the ground state energy of our model
 in finite volume, so that at the end we can glue it with the previous set of equations into a single set of 
 two-component nonlinear integral equations, 
 as it was done earlier in the $S \to \infty$ limit of the isotropic spin-S vertex model \cite{HA}.
  Therefore let us formulate our RSOS($k,q$) model as an RSOS($k',q'$) model with $k'=r-2-k$,
 $q'=r-2-q$. In this case the T-system (\ref{16}) is of the form:  
 \begin{equation} 
\hat{T}^{(k')}_p(x+i)\hat{T}^{(k')}_p(x-i)=\hat{f}^{(k')}_p (x)+\hat{T}^{(k')}_{p-1}(x)\hat{T}^{(k')}_{p+1}(x), \label{72}
\end{equation}
where
\begin{equation}
\hat{f}^{(k')}_p(x)=\hat{T}^{(k')}_0(x+i(p+1)) \ \hat{T}^{(k')}_0(x-i(p+1)),
\end{equation}  
and
\begin{equation}
\hat{T}^{(k')}_{-1}(x)=0, \qquad \qquad \hat{T}^{(k')}_0(x)=\prod_{j=0}^{k'-1} \phi(x+i(k'-1-2j)).
\end{equation}
The solutions of these equations are of the form
\begin{equation}
\hat{T}^{(k')}_p(x)=\sum_{l=1}^{p+1} \hat{\lambda}_{l}^{(p)}(x),
\end{equation}
where
\begin{equation}
\hat{\lambda}_{l}^{(p)}(x)=  
\hat{T}^{(k')}_0(x+i(2l-p-2)) \ \frac{e^{i \omega (p+2-2l)} \ Q^{(k')}(x+i(p+1)) \ Q^{(k')}(x-i(p+1)) }
{Q^{(k')}(x+i(2l-p-1)) \ Q^{(k')}(x+i(2l-p-3)) }. \label{76}
\end{equation}
We define the following auxiliary functions: 
\begin{equation}
y^{(k')}_j(x)=\frac{\hat{T}^{(k')}_{j-1}(x) \hat{T}^{(k')}_{j+1}(x) }{\hat{f}^{(k')}_j(x)}, \ \qquad \qquad j=1,\dots,k'+1, \label{77}
\end{equation}
\begin{equation}
Y^{(k')}_j(x)=1+y^{(k')}_j(x)=\frac{ \hat{T}^{(k')}_j (x+i) \hat{T}^{(k')}_j (x-i) }{\hat{ f}^{(k')}_j(x) }, 
\ \qquad \qquad j=1,\dots,k'+1, \label{78}
\end{equation}
\begin{equation}
h(x)=\frac{ \hat{\lambda}_{1}^{(k'+1)}(x+i)+\dots+\hat{\lambda}_{k'+1}^{(k'+1)}(x+i) }{\hat{\lambda}_{k'+2}^{(k'+1)}(x+i) },
\qquad \qquad \mathcal{H}(x)=1+h(x), \label{79}
\end{equation}
\begin{equation}
\bar{h}(x)=\frac{ \hat{\lambda}_{2}^{(k'+1)}(x-i)+\dots+\hat{\lambda}_{k'+2}^{(k'+1)}(x-i) }{\hat{\lambda}_{1}^{(k'+1)}(x-i) },
\qquad \qquad \bar{\mathcal{H}}(x)=1+\bar{h}(x). \label{80}
\end{equation}
From (\ref{72}-\ref{76}) it follows that the auxiliary functions (\ref{77}-\ref{80}) are periodic functions with period $2ir$
and satisfy the following functional relations:
\begin{equation}
\hat{T}^{(k')}_{k'+1}(x+i)=e^{-i \omega (k'+1)} \ \hat{T}^{(k')}_0(x+i(k'+2)) \ \frac{Q^{(k')}(x-ik'-i) }{ Q^{(k')}(x+ik'+i) } \
\mathcal{H}(x), \label{81}
\end{equation}
\begin{equation}
\hat{T}^{(k')}_{k'+1}(x-i)=e^{i \omega (k'+1)} \ \hat{T}^{(k')}_0(x-i(k'+2)) \ \frac{Q^{(k')}(x+ik'+i) }{ Q^{(k')}(x-ik'-i) } \
\bar{\mathcal{H}}(x),\label{82}
\end{equation}
\begin{equation}
h(x)=\frac{e^{i \omega (k'+2)}}{\hat{T}^{(k')}_0(x+i(k'+2))} \  \frac{ Q^{(k')}(x+ik'+3i) }{ Q^{(k')}(x-ik'-i) } \ \hat{T}^{(k')}_{k'}(x),
\end{equation}
\begin{equation}
\bar{h}(x)= \frac{e^{-i \omega (k'+2)}}{\hat{T}_0(x-i(k+2))}\  \frac{ Q^{(k')}(x-ik'-3i) }{ Q^{(k')}(x+ik+i) } \ \hat{T}^{(k')}_{k'}(x),
\end{equation}
\begin{equation}
\mathcal{H}(x) \bar{\mathcal{H} }(x)=Y^{(k')}_{k'+1}(x), \label{85}
\end{equation}
\begin{equation}
y^{(k')}_j(x+i) y^{(k')}_j(x-i)=Y^{(k')}_{j-1}(x) Y^{(k')}_{j+1}(x) \qquad \qquad j=1,\dots,k'-1,
\end{equation}
\begin{equation}
y^{(k')}_{k'}(x+i) y^{(k')}_{k'}(x+i) = Y^{(k')}_{k'-1}(x) \mathcal{H}(x) \bar{\mathcal{H} }(x),
\end{equation}
\begin{equation}
\hat{T}^{(k')}_{k'}(x+i) \hat{T}^{(k')}_{k'}(x-i)=\hat{T}^{(k')}_0(x+i(k'+1)) \ \hat{T}^{(k')}_0(x-i(k'+1))  Y^{(k')}_{k'}(x). \label{88}
\end{equation}
In order to be able to derive integral equations from these functional relations one needs to know the positions of the
zeroes and the poles of the auxiliary functions (\ref{77}-\ref{80}). Due to the relations (\ref{81}-\ref{88}) we only need
  the zeroes of 
$Q^{(k')}(x)$ and $\hat{T}^{(k')}_j(x)$. The zeroes of $Q^{(k')}(x)$ are the solutions of the Bethe Ansatz equations (\ref{24}-\ref{26})
with the $k\to k'$ change. Therefore the zeroes of $Q^{(k')}(x)$ form $N/2$ pieces of $k'$-strings. As it was mentioned in the previous
section there are deviations from the prediction of string hypothesis, but in the imaginary direction
these deviations are always less than $1/2$. 

As far as the zeroes of the $\hat{T}^{(k')}_j(x)$ transfer matrices are concerned,
 all $\hat{T}^{(k')}_j(x)$ can have  $\frac{N}{2}$-fold degenerate 
zeroes in  the \lq\lq main" strip at places $\pm x_0, \pm x_0 \pm i$, but these zeroes cancel from the Y-system 
elements (\ref{77}-\ref{78}) except the $y^{(k')}_{k'}(x)$ case which has $\frac{N}{2}$-fold degenerate zeroes in the 
\lq\lq main" strip at $\pm x_0$.
 These zeroes will give the standard TBA source term in our final equations \cite{Kuniba,SGg}.
Let us introduce
\begin{equation}
\Psi^{(k')}_1(x)=Q^{(k')}(x-ik')
\end{equation}
\begin{equation}
\Psi^{(k')}_2(x)=Q^{(k')}(x+ik')=\Psi^{(k')}_1(x-2i(r-k'))
\end{equation}
 Now we can list the strips,
where the auxiliary functions are analytic and non zero (ANZ): 
\begin{eqnarray}
\Psi^{(k')}_1(x) \qquad  &\mbox{ANZ}& \qquad -2(r-k')+1/2 \leq  \mbox{Im}\ x \leq 1/2, \ \ \mbox{mod}  \ 2r  \nonumber  \\ 
\Psi^{(k')}_2(x) \qquad &\mbox{ANZ}& \qquad -1/2 \leq \mbox{Im}\ x \leq 2(r-k')+1/2,   \ \ \mbox{mod} \ 2r  \nonumber  \\
h(x),\mathcal{H}(x) \qquad &\mbox{ANZ}& \qquad -3/2 < \mbox{Im}\ x \leq 0,  \ \  \mbox{mod} \ 2r \nonumber  \\
\bar{h}(x),\bar{\mathcal{H}}(x) \qquad &\mbox{ANZ}& \qquad 0 \leq \mbox{Im}\ x < 3/2,  \ \ \mbox{mod}  \ 2r \label{}\\
y^{(k')}_j(x)  \qquad &\mbox{ANZ}& \qquad 0 \leq |\mbox{Im}\ x| \leq 1,  \ \ \mbox{mod} \ 2r \qquad j=1,\dots,k-1,    \nonumber  \\ 
Y^{(k')}_j(x) \qquad &\mbox{ANZ}& \qquad 0 \leq |\mbox{Im}\ x| \leq \epsilon,   \ \ \mbox{mod} \ 2r \qquad j=j,\dots,k \quad \epsilon>0,   
\nonumber  \\
\hat{T}^{(k')}_{k'+1}(x)  \qquad &\mbox{ANZ}& \qquad 0 \leq |\mbox{Im}\ x| \leq 1  \ \ \mbox{mod} \ 2r. \nonumber 
\end{eqnarray}
We introduce new variables by shifting the arguments of $h(x),\mathcal{H}(x)$ and $\bar{h}(x),\bar{\mathcal{H}}(x)$ by $\pm i \gamma'$:  
\begin{equation}
a(x)=h(x-i\gamma'), \qquad \qquad U(x)=\mathcal{H}(x-i\gamma')=1+a(x),
\end{equation}
 \begin{equation}
\bar{a}(x)=\bar{\mathcal{H}}(x+i\gamma'), \qquad \qquad \bar{U}(x)=\bar{\mathcal{H}}(x+i\gamma')=1+\bar{a}(x),
\end{equation}
where $0<\gamma'<1/2$ is an arbitrary real and fixed parameter. This shift  is necessary because the original functions
$\mathcal{H}(x),\bar{\mathcal{H}}(x)$ have zeroes and poles close to the real axis on the upper and lower half plane respectively.
Due to the ANZ property of  $\hat{T}^{(k')}_{k'+1}(x)$ (\ref{}) and the fact that $\lim_{|x| \to \infty} \frac{d^2}{dx^2} \log 
 \hat{T}^{(k')}_{k'+1}(x) = 0$ the following relation holds 
\begin{equation}
0=\int\limits_{-\infty}^{\infty}dx \ \frac{d^2}{dx^2} \log  \hat{T}^{(k')}_{k'+1}(x-i)\ e^{iq(x-i)}-\int\limits_{-\infty}^{\infty} dx \
\frac{d^2}{dx^2} \log  \hat{T}^{(k')}_{k'+1}(x+i)\ e^{iq(x+i)}. \label{94}
\end{equation}
From (\ref{81}) and (\ref{82}) one can express $\hat{T}^{(k')}_{k'+1}(x \pm i)$ with the auxiliary functions and by substituting these 
expressions  into (\ref{94}) one gets  in  Fourier space 

\begin{eqnarray}
\tilde{d^2 l} \Psi^{(k')}_1 (q) 
= \frac{   e^{(r-k')q} \ \tilde{d^2 l} f(q) +e^{(r-k'+1-\gamma')q } \ \ \tilde{d^2 l} \bar{U} (q) - 
 e^{(r-k-1+\gamma')q } \ \ \tilde{d^2 l} {U} (q) }{4 \ \cosh (q) \ \sinh[(r-k')q] },  \\
\tilde{d^2 l} \Psi^{(k')}_2 (q) 
= \frac{   e^{-(r-k')q} \ \tilde{d^2 l} f(q) +e^{-(r-k'+1+\gamma')q } \ \ \tilde{d^2 l} \bar{U} (q) - 
 e^{-(r-k'-1-\gamma')q } \ \ \tilde{d^2 l} {U} (q) }{4 \ \cosh (q) \ \sinh[(r-k')q] },  \\
 \end{eqnarray}
where 
\begin{equation}
\tilde{d^2l} f(q)=\tilde{d^2 l} \left( \hat{T}^{(k')}_0 \right)^{-(k'+1)}(q)-\tilde{d^2 l} 
\left( \hat{T}^{(k')}_0 \right)^{+(k'+1)}(q).
\end{equation}
After a similar procedure that has been done in the previous section one gets the following nonlinear integral equations in the 
continuum  
\begin{eqnarray}
\log y^{(k')}_1(x) &=& (K*\log Y^{(k')}_2)(x),  \nonumber  \\
\log y^{(k')}_j (x) &=& (K* \log Y^{(k')}_{j-1})+(K*\log Y^{(k')}_{j+1})(x), \qquad  \quad j=2,\dots,k'-1   \nonumber  \\
\log y^{(k')}_{k'}(x) &=& {\mathcal{D}}^{TBA}_N(x)+ (K*\log Y^{(k')}_{k'-1})(x)+ (K^{+\gamma'}*\log U)(x) \nonumber  \\
                        &+& (K^{-\gamma'}*\log \bar{U})(x)- \delta \pi, \label{99}  \\
\log a(x) &=&  (\tilde{G}*\log  U)(x)- (\tilde{G}^{+2(1-\gamma')}*\log \bar{U} )(x) 
                            + (K^{-\gamma'}*\log Y^{(k')}_{k'})(x)+\frac{i \ \pi}{r-k'-1},  \nonumber  \\
\log \bar{a}(x) &=&  (\tilde{G}*\log  \bar{U})(x)- (\tilde{G}^{-2(1-\gamma')}*\log  U )(x)  
                             + (K^{+\gamma'}*\log Y^{(k')}_{k'})(x)-\frac{i \ \pi}{r-k'-1}, \nonumber 
\end{eqnarray}
where the \lq\lq source" function on the lattice reads as
\begin{equation}
{\mathcal{D}}^{TBA}_N(x)=\log \left((-1)^{\delta} \ \tanh^{N/2}\left( \frac{\pi}{4}(x-x_0)\right) \ 
\tanh^{N/2}\left( \frac{\pi}{4}(x+x_0)\right) \right)
\end{equation}
which becomes the usual TBA source term in the continuum
\begin{equation}
\lim_{N \to \infty} {\mathcal{D}}^{TBA}_N(x)= -ml \cosh\left( \frac{\pi}{2} x \right).
\end{equation}
The kernel function $K(x)$ and $\delta$ are the same as in the previous section and
\begin{equation}
\tilde{G}(x)=\int\limits_{-\infty}^{\infty} \frac{dq}{2 \pi} \ \frac{ e^{iqx} \ \sinh[(r-k'-2)q]}{2 \ \cosh(q) \ \sinh[(r-k'-1)q] }
=\int\limits_{-\infty}^{\infty} \frac{dq}{2 \pi} \ \frac{e^{iqx} \ \sinh(kq)}{2 \ \cosh(q) \ \sinh[(k+1)q] }, \label{100}
\end{equation}
The correct choice of the value of $\delta$ is zero for the ground state of the $\cal{M}$$A^{+}_{k,r}$.
In the TBA language the complex auxiliary functions $a(x)$ and $\bar{a}(x)$ resum the contributions of 
those Y-system elements whose index is larger than $k'$. (See figure \ref{2f}$b$.)



Due to the identity $Y^{(k)}_{k}(x)=Y^{(k')}_{k'}(x)$, which follows from (\ref{35})
the energy (\ref{6}) of the continuum model can be expressed by a gauge invariant Y-system element of the RSOS($k',q'$) model:
\begin{equation}
E_0(l)=E_{bulk}-\frac{m}{4} \int\limits_{-\infty}^{\infty} dx  \cosh \left( \frac{\pi x}{2}\right) \log Y^{(k')}_{k'}(x). \label{gaugeY}
\end{equation}
Equations (\ref{62}) and (\ref{99}) are different descriptions of  the ground
state energy of the massive $\phi_{\mbox{id,id,adj}}$ perturbation of the
$SU(2)_k \times SU(2)_{k'} /SU(2)_{k+k'}$ model.
 In this paper one of our main purposes is to describe the finite size 
dependence of the ground state energy of this model with a two-component nonlinear integral equations. 
As we will see in the next section following the idea of \cite{HA}
we are able to construct such two component nonlinear integral equations by combining
 (\ref{62}) and (\ref{99}).


\section{The two component nonlinear integral equations}

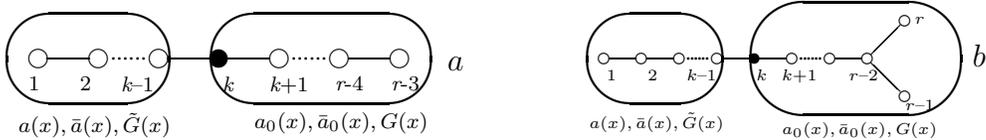
\begin{figure}[htbp]
\begin{center}
\begin{picture}(280,30)(-140,-15)
\put(-163,-7) {\usebox{\RSN}}
\put(-74,-7) {\usebox{\FSSGN}}
\put(-130,-15){\parbox{130mm}{\caption{ \label{3f}\protect {\small
Graphical representation of the two component nonlinear integral equations of the $a$.) $\cal{M}$$A^{+}_{k,r}$ and
$b$.) FSSG model }}}}
\end{picture}
\end{center}
\end{figure}

In this section we construct two component nonlinear integral equations from (\ref{62}) and (\ref{99}).
 The complex auxiliary functions (\ref{44}-\ref{45}) and (\ref{79}-\ref{80}) 
 are gauge dependent and have a finite continuum limit only in an appropriate gauge, but the Y-system 
elements (\ref{29}-\ref{31}) are gauge invariant and  have finite continuum limit. One can see from (\ref{50}) and (\ref{85})
 that some gauge 
invariant Y-system elements can be expressed by these non-gauge invariant auxiliary functions:
\begin{equation}
U_0(x+i \gamma) \bar{U}_0(x-i \gamma)=Y^{(k)}_k(x), \label{}
\end{equation} 
\begin{equation}
U(x+i \gamma') \bar{U}(x-i \gamma')=Y^{(k')}_{k'+1}(x).
\end{equation}
In our case because of the equivalence of the Y-systems of an RSOS($k,q$) and RSOS($k',q'$) model (\ref{35})
the following relation will also be true
\begin{equation}
U_0(x+i \gamma) \bar{U}_0(x-i \gamma)=Y^{(k')}_{k'}(x). \label{104a}
\end{equation}
\begin{equation}
U(x+i \gamma') \bar{U}(x-i \gamma')=Y^{(k)}_{k-1}(x). \label{104}
\end{equation}
Substituting (\ref{104}) into the equations for $a_0(x), \bar{a}_0(x)$ in  (\ref{62}) and  (\ref{104a}) into the equations for  
$a(x),\bar{a}(x)$ in  (\ref{99}) and deforming  the integration contour appropriately, we get the two component nonlinear
 integral equations for the ground state of our model ($\delta \equiv 0$).
The equations are as follows:
\begin{eqnarray}
\log a_0(x) &=& \mathcal{D}(x-i \gamma)+ (G*\log  U_0)(x)- (G^{+2(1-\gamma)}*\log \bar{U}_0 )(x) \nonumber \\    
                            &+&  (K^{+(\gamma'-\gamma)}*\log U )(x)+ (K^{-(\gamma'+\gamma)}*\log \bar{U} )(x)
			    +i\frac{\pi}{r-k},
 \nonumber  \\
\log \bar{a}_0(x) &=& \mathcal{D}(x+i \gamma)+ (G*\log  \bar{U}_0)(x)- (G^{-2(1-\gamma)}*\log {U}_0 )(x)     \nonumber  \\
                            &+&  (K^{+(\gamma'+\gamma)}*\log U )(x)+ (K^{+(\gamma-\gamma')}*\log \bar{U} )(x)-i\frac{\pi}{r-k},
 \nonumber  \\
\log a(x) &=&  (\tilde{G}*\log  U)(x)- (\tilde{G}^{+2(1-\gamma')}*\log \bar{U} )(x)  \label{105} \\ 
                            &+&  (K^{+(\gamma-\gamma')}*\log U_0 )(x)+  (K^{-(\gamma'+\gamma)}*\log \bar{U}_0 )(x)+i\frac{\pi}{k+1},
 \nonumber  \\
\log \bar{a}(x) &=&  (\tilde{G}*\log \bar{U})(x)- (\tilde{G}^{-2(1-\gamma')}*\log {U} )(x)     \nonumber \\ 
                            &+&  (K^{+(\gamma+\gamma')}*\log U_0 )(x)+ (K^{+(\gamma'-\gamma)}*\log \bar{U}_0 )(x)-i\frac{\pi}{k+1},
 \nonumber  \\
U_0(x)&=&1+a_0(x), \enskip \bar{U}_0(x)=1+\bar{a}_0(x), \enskip U(x)=1+a(x), \enskip
\bar{U}(x)=1+\bar{a}(x). \nonumber 
\end{eqnarray}
where $0<\gamma \leq 1/2$, $0<\gamma'\leq 1/2$ are arbitrary fixed real parameters.
 Equations (\ref{105}) are graphically represented in figure \ref{3f}$a$.
 
The energy expression is the same as (\ref{68}). This is a closed set of equations for four complex unknown functions
\lq\lq $a(x),\bar{a}(x),a_0(x),\bar{a}_0(x)$", but $\bar{a}(x)$  is the complex conjugate of $a(x)$ and  $\bar{a}_0(x)$  
is the complex conjugate of $a_0(x)$, therefore only four real unknown functions describe the model.

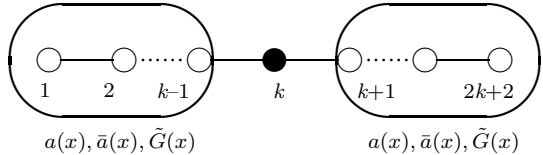
\begin{figure}[htbp]
\begin{center}
\begin{picture}(280,30)(-140,-15)
\put(-110,-10) {\usebox{\NSYM}}

\put(-120,-19){\parbox{130mm}{\caption{ \label{4f}\protect {\small
Graphical representation of the NLIEs of the $\cal{M}$$A^{+}_{k,2k+2}$ model. }}}}
\end{picture}
\end{center}
\end{figure}

For the special case of $r=2k+2$,
from (\ref{99}) one can derive a different set of nonlinear integral equations to describe the finite size dependence of the 
ground  state energy of our model. In this case $k=k'$ and thus $Y^{(k')}_{k'-1}(x)=Y^{(k)}_{k-1}(x)=U(x+i\gamma') U(x-i\gamma')$.
Replacing this into the equation of $y^{(k')}_{k'}(x)$ in (\ref{99})

and deforming the integration contour appropriately 
we get the following equations
 \begin{eqnarray}
\log y_{0}(x) &=& \mathcal{D}(x)+2 (K^{+\gamma'}*\log U)(x)+
                                         2(K^{-\gamma'}*\log \bar{U})(x), \nonumber  \\
\log a(x) &=&  (\tilde{G}*\log  U)(x)- (\tilde{G}^{+2(1-\gamma')}*\log \bar{U} )(x)  \nonumber \\
               &+&  (K^{-\gamma'}*\log Y_{0})(x)+\frac{i \pi}{k+1}, \label{106}   \\
\log \bar{a}(x) &=&  (\tilde{G}*\log  \bar{U})(x)- (\tilde{G}^{-2(1-\gamma')}*\log  U )(x) \nonumber \\  
                          &+&  (K^{+\gamma'}*\log Y_{0})(x)-\frac{i \pi}{k+1}, \nonumber \\
 U(x)&=&1+a(x), \quad \bar{U}(x)=1+\bar{a}(x), \nonumber 
\end{eqnarray}
where we introduced the notations:
\begin{equation}
y_0(x)=y^{(k)}_k(x), \qquad \qquad Y_0(x)=Y^{(k)}_k(x).
\end{equation}
Equations (\ref{106}) are graphically represented in figure \ref{4f}. 

In this case the ground state energy of the model  is of the form
\begin{equation}
E_0(l)=E_{bulk}-\frac{m}{4} \int\limits_{-\infty}^{\infty} dx  \cosh \left( \frac{\pi x}{2}\right) \log Y_0(x).
\end{equation}
 Equations (\ref{106}) contain only three real unknown functions because $y_0(x)$ and $Y_0(x)$ are real and $\bar{a}(x)$ 
 is the complex  conjugate of  $a(x)$. 

It can be easily seen that the $k \to \infty$ limit of equations (\ref{106}) are the same as the ones which were proposed
earlier \cite{HA} for the $O(4)$ NLS model.
For finite $k$ these equations describe the 
$\cal{M}$$A^{+}_{k,2k+2}$ model, the $k \to \infty$ limit of which is the
$O(4)$ nonlinear $\sigma$-model \cite{fon}.

\section{The test of  the equations }

In this section we will make some analytical and numerical tests on our equations (\ref{105}). First we calculate the UV central charge
of the model using the equations (\ref{105}). Using the standard method of  refs. \cite{KBP,ZamiC} the energy in the conformal limit ($l \rightarrow 0$) 
can be expressed by the dilogarithm functions. The energy in the conformal limit is of the form:
\begin{eqnarray}
E_0(l) \simeq \frac{1}{\pi l} \bigg( L_{+}(a_{+}(\infty))+L_{+}(\bar{a}_{+}(\infty))+L_{+}(a_{0+}(\infty))+L_{+}(\bar{a}_{0+}(\infty))
 \nonumber    \\  
-L_{+}(a_{+}(-\infty))-L_{+}(\bar{a}_{+}(-\infty))-L_{+}(a_{0+}(-\infty))-L_{+}(\bar{a}_{0+}(-\infty)) \bigg) \label{109} \\ \nonumber
+i\frac{\pi}{2(r-k)} \left\{ \left( lA_{0+}(+\infty)-\bar{lA}_{0+}(+\infty) \right)-\left(lA_{0+}(-\infty)-\bar{lA}_{0+}(-\infty) 
\right) \right\} \\ \nonumber
+i\frac{\pi}{2(k+1)} \left\{ \left( lA_{+}(+\infty)-\bar{lA}_{+}(+\infty) \right)-\left( lA_{+}(-\infty)-\bar{lA}_{+}(-\infty) 
\right) \right\}
\end{eqnarray}
where 
\begin{eqnarray}
lA_{+}(x) &=& \log(1+a_{+}(x)), \quad \ \bar{lA}_{+}(x)=\log(1+\bar{a}_{+}(x)), \\ \nonumber
lA_{0+}(x) &=& \log(1+a_{0+}(x)) \quad \  \bar{lA}_{0+}(x)=\log(1+\bar{a}_{0+}(x)),
\end{eqnarray}
$L_{+}(z)$ is defined by the integral
\begin{equation}
L_{+}(z)=\frac12 \int\limits_{0}^{z} dx \left( \frac{\log(1+x)}{x}-\frac{x}{1+x} \right), \qquad L_{+}(x)=L \left(\frac{x}{1+x}\right),
\end{equation}
and $L(x)$ is Roger's dilogarithm function. The functions $a_{+}(x),\bar{a}_{+}(x),a_{0+}(x),\bar{a}_{0+}(x)$ 
denote the kink functions corresponding to $a(x),\bar{a}(x),a_{0}(x),\bar{a}_{0}(x)$ respectively.
 The limits of these kink functions at infinity are as follows:
\begin{equation}
a_{+}(\infty)=(\bar{a}_{+}(\infty))^{*}=e^{i \frac{2\pi}{k+2}},
 \quad a_{+}(-\infty)=(\bar{a}_{+}(-\infty))^{*}=e^{i \frac{(k'+2)\pi}{r}} 
 \frac{\sin\left(\frac{(k'+1) \pi}{r} \right) }{\sin \left(\frac{\pi}{r} \right)},
\end{equation}
\begin{equation}
a_{0+}(\infty)=\bar{a}_{0+}(\infty)=0, 
\quad a_{0+}(-\infty)=(\bar{a}_{0+}(-\infty))^{*}=e^{i \frac{(k+1)\pi}{r}} 
 \frac{\sin\left(\frac{k \pi}{r} \right) }{\sin \left(\frac{\pi}{r} \right)},
\end{equation}
where the * denotes the complex conjugation.
From these using some dilogarithm identities \cite{kirill}, 
one gets the ground state energy in the $l \rightarrow 0$ limit
\begin{equation}
E_{0}(l) \simeq -\frac{c_{UV}(k,r) \pi}{ 6 \ l}.
\end{equation} 
Where $c_{UV}(k,r)$ is the effective UV central charge of the model, the value of which according to our equations is 
\begin{equation}
c_{UV}(k,r)=\frac{3k}{k+2} \left( 1-\frac{2\ (k+2)}{r \ (r-k)}\right)
\end{equation}
which agrees with the UV central charge of the $\cal{M}$$A^{+}_{k,r}$ model.
So we have checked our equations (\ref{105}) analytically in the small $l$ limit in leading order. 
Next we will check our equations in the large $l$ limit analytically in leading order.
The equations (\ref{105}) can be solved iteratively in the large $l$ regime. After some easy calculations one gets that the
 ground state energy is of the form
 \begin{equation}
 E_0(l)=E^{(1)}+O(e^{-2ml}),  \qquad  \label{117}
 \end{equation} 
 where
 \begin{equation}
 E^{(1)}=-4\ \cos\left( \frac{\pi}{r-k} \right) \ \cos\left( \frac{\pi}{k+2} \right)
  \int\limits_{-\infty}^\infty d\theta \  \frac{m}{2 \pi} \ \cosh(\theta) \ e^{-ml \cosh(\theta)},  \label{117a}
 \end{equation}
 which agrees perfectly with TBA results obtained for $\cal{M}$$A^{+}_{k,r}$ models \cite{FSSS}.
 Note, that this expression is in accordance with the expectation for the low-temperature
 behaviour of the kink system with the vacuum structure described in the introduction.
 In (\ref{117a}) the factor in front of the integral expression takes into account the statistics 
 of interkink "colouring" with the $A_{k+1} \otimes A_{k'+1}$ adjacency structure. Therefore this limit
 is in agreement with the $(\mbox{RSOS})_k \times (\mbox{RSOS})_{k'}$ scattering theory of ref.\cite{ABL}.
 
So far we have made some analytical tests on our equations (\ref{105}) in the small $l$ and large $l$ regimes.
We also made numerical calculations to check the correctness of our equations (\ref{105}).
Therefore for different values of the parameters $k,r$ and $l$, we solved numerically our eqs. (\ref{105}),
the corresponding TBA equations \cite{FSSS}, and Dunning's \lq\lq half TBA-NLIE" type of equations \cite{dun}.
In every case all the three type of equations gave the same numerical results, which convinced us about
the correctness of our equations (\ref{105}).


\section{Two component nonlinear integral equations for the fractional supersymmetric sine-Gordon models}

In this section we propose two component nonlinear integral equations for the fractional supersymmetric sine-Gordon models. 
One can recognize that taking the $r \to \infty$ limit of eqs. (105), one obtains two component nonlinear 
integral equations for the current-current perturbation of the $SU(2)_k$ WZW  model, which is a fractional 
supersymmetric sine-Gordon model at a special value of the coupling. The equations take the form
\begin{eqnarray}
\log a_0(x) &=& \mathcal{D}(x-i \gamma)+ (F*\log  U_0)(x)- (F^{+2(1-\gamma)}*\log \bar{U}_0 )(x) \nonumber \\    
                            &+&  (K^{+(\gamma'-\gamma)}*\log U )(x)+ (K^{-(\gamma'+\gamma)}*\log \bar{U} )(x),
 \nonumber  \\
\log \bar{a}_0(x) &=& \mathcal{D}(x+i \gamma)+ (F*\log  \bar{U}_0)(x)- (F^{-2(1-\gamma)}*\log {U}_0 )(x)     \nonumber  \\
                            &+&  (K^{+(\gamma'+\gamma)}*\log U )(x)+ (K^{+(\gamma-\gamma')}*\log \bar{U} )(x),
 \nonumber  \\
\log a(x) &=&  (\tilde{G}*\log  U)(x)- (\tilde{G}^{+2(1-\gamma')}*\log \bar{U} )(x)  \label{119} \\ 
                            &+&  (K^{+(\gamma-\gamma')}*\log U_0 )(x)+  (K^{-(\gamma'+\gamma)}*\log \bar{U}_0 )(x)+i\frac{\pi}{k+1},
 \nonumber  \\
\log \bar{a}(x) &=&  (\tilde{G}*\log \bar{U})(x)- (\tilde{G}^{-2(1-\gamma')}*\log {U} )(x)     \nonumber \\ 
                            &+&  (K^{+(\gamma+\gamma')}*\log U_0 )(x)+ (K^{+(\gamma'-\gamma)}*\log \bar{U}_0 )(x)-i\frac{\pi}{k+1},
 \nonumber  \\
U_0(x)&=&1+a_0(x), \enskip \bar{U}_0(x)=1+\bar{a}_0(x), \enskip U(x)=1+a(x), \enskip
\bar{U}(x)=1+\bar{a}(x). \nonumber 
\end{eqnarray}
where where $0<\gamma \leq 1/2$, $0<\gamma'\leq 1/2$ are arbitrary fixed real parameters, 
$\tilde{G}(x)$ is the same as in (100), but
\begin{equation}
F(x)=\int\limits_{-\infty}^{\infty} \frac{dq}{2 \pi} \  \frac{e^{-|q|-iqx}}{2 \cosh(q) },
\end{equation}
is the $r \to \infty$ limit of $G(x)$, and the energy formula agrees with (\ref{68}).
 One can see in eqs. (\ref{119}), that the variables $a(x),\bar{a}(x)$ resum the
contributions of those nodes of the corresponding Dynkin-diagram, whose index are less than $k$. 
The lattice version of these equations correspond to the case of the isotropic spin-$k/2$ vertex model. 
So far, all of the two component nonlinear integral equations we produced, were the results of derivations.
Now we will propose two component nonlinear integral equations for the fractional supersymmetric sine-Gordon models,
 the lattice version of which will correspond to the case of the anisotropic higher spin vertex model.

One can see in eqs. (\ref{119})
that those components of the equations, which have $\log a(x)$ and $\log \bar{a}(x)$ on the left hand side, are independent
of $r$ (the entire length of the Dynkin-diagram) and 
the only parameter which appears in them is the $k$, the length of the resummed part of the Dynkin diagram.
From this one can conclude, that the form of these two components of the equations remain the same, if the right hand side of the
Dynkin-diagram is changed from $A_{r-3}$ to $D_r$. Therefore deforming the kernel function $F(x)$ of eqs. (\ref{119}) 
to $G(x)$, as it was done in \cite{dun},
 we obtain two component nonlinear integral equations for the FSSG models.
  The proposed equations are as follows
\begin{eqnarray}
\log a_0(x) &=& \mathcal{D}(x-i \gamma)+ (G*\log  U_0)(x)- (G^{+2(1-\gamma)}*\log \bar{U}_0 )(x) \nonumber \\    
                            &+&  (K^{+(\gamma'-\gamma)}*\log U )(x)+ (K^{-(\gamma'+\gamma)}*\log \bar{U} )(x),
 \nonumber  \\
\log \bar{a}_0(x) &=& \mathcal{D}(x+i \gamma)+ (G*\log  \bar{U}_0)(x)- (G^{-2(1-\gamma)}*\log {U}_0 )(x)     \nonumber  \\
                            &+&  (K^{+(\gamma'+\gamma)}*\log U )(x)+ (K^{+(\gamma-\gamma')}*\log \bar{U} )(x),
 \nonumber  \\
\log a(x) &=&  (\tilde{G}*\log  U)(x)- (\tilde{G}^{+2(1-\gamma')}*\log \bar{U} )(x)  \label{121} \\ 
                            &+&  (K^{+(\gamma-\gamma')}*\log U_0 )(x)+  (K^{-(\gamma'+\gamma)}*\log \bar{U}_0 )(x)+i\frac{\pi}{k+1},
 \nonumber  \\
\log \bar{a}(x) &=&  (\tilde{G}*\log \bar{U})(x)- (\tilde{G}^{-2(1-\gamma')}*\log {U} )(x)     \nonumber \\ 
                            &+&  (K^{+(\gamma+\gamma')}*\log U_0 )(x)+ (K^{+(\gamma'-\gamma)}*\log \bar{U}_0 )(x)-i\frac{\pi}{k+1},
 \nonumber  \\
U_0(x)&=&1+a_0(x), \enskip \bar{U}_0(x)=1+\bar{a}_0(x), \enskip U(x)=1+a(x), \enskip
\bar{U}(x)=1+\bar{a}(x). \nonumber 
\end{eqnarray}
where kernel functions are the same as in (\ref{105}) and the energy expression is also the same (\ref{68}). 
These equations describe the ground state energy of the fractional supersymmetric sine-Gordon models, where the
relations between the $\beta^2$ coupling of the model, and the parameters $r$ and $k$ of the eqs. (\ref{121}) are
\begin{equation}
\frac{\beta^2}{8\pi}=\frac{r-k}{r \ k}.
\end{equation}
For the graphical representation of the equations (\ref{121}) see figure \ref{3f}$b$.
From the eqs. (\ref{121}) after some usual straightforward calculations one can easily obtain the expected value of 
the central charge $c_{UV}=\frac{3k}{k+2}$, and for the leading order infrared behavior of the energy one 
obtains 
 \begin{equation}
 E_0(l) \sim -4 \ \cos\left( \frac{\pi}{k+2} \right)
 \frac{m}{2 \pi} \int\limits_{-\infty}^\infty d\theta \ \cosh(\theta) \ e^{-ml \cosh(\theta)},  
 \end{equation}
which agrees perfectly with TBA results.
 
 We mention, that eqs. (\ref{121}) could be obtained by the following formal train of thought:
 the FSSG models were first proposed as the models obtained by \lq\lq unrestricting" the S-matrices
 describing  $\cal{M}$$A^{+}_{k,r}$ models \cite{BL}.
 Eqs. (\ref{105}) describe the finite size effects in this model, 
 and so we can formally \lq\lq release" the restriction by removing the $r$ dependent imaginary constant terms from (\ref{105}), 
 and thus we obtain eqs. (\ref{121}). Then by formally \lq\lq releasing" the remaining restriction in eqs. (\ref{121})
 by removing the remaining $k$-dependent imaginary constant, one obtains
 the earlier proposed two component nonlinear integral equations of the SS-model \cite{HA}.

We also made numerical calculations to check the correctness of our conjectured equations (\ref{121}).
Therefore for different values of the parameters $k,r$ and $l$, we solved numerically our eqs. (\ref{121}),
the corresponding TBA equations \cite{fi}, and the Dunning's \lq\lq half TBA-NLIE" type of equations \cite{dun}.
In every case all the three type of equations gave the same numerical results, which convinced us about
the correctness of our equations (\ref{121}).


\section{The equations for finite solvable lattices}

In this section we briefly summarize how the two component nonlinear integral equations
of the previous sections have to be modified for the case of solvable lattice models.
In the context of lattice models our purpose is to calculate exactly the eigenvalues of the
transfer matrices (\ref{14}) (with $x_0=0$) for finite $N$, because by taking the first derivative of these 
transfer matrices at certain points one can define integrable spin Hamiltonians from them \cite{BZR,KR}.
 The derivation described in the previous sections can be achieved 
for the largest eigenvalue of the transfer matrix of the critical RSOS($k,q$) model.
The equations are the same as (\ref{105}) with the change of the source function $\mathcal{D}(x)$ to
\begin{equation}
\hat{\mathcal{D}}_N (x)= i N \arctan \left[  \sinh \left(  \frac{\pi (x+i)}{2}   \right)    \right] .
\end{equation}
The largest eigenvalue of the transfer matrix  is given by
\begin{eqnarray}
\log \Lambda^{(k)}_k (x,N) &=& \log \Lambda^{(k)}_{k,bulk}(x,N)+\frac{1}{4} \int\limits_{-\infty}^{\infty} dx'  \ 
 \frac{\log U_0(x')}{\cosh \left( \frac{\pi (x-x'+i \gamma)}{2}\right) } \label{125} \\
&+& \frac{1}{4} \int\limits_{-\infty}^{\infty} dx' \ 
\frac{\log \bar{U}_0(x')}{\cosh \left( \frac{\pi (x-x'-i \gamma)}{2}\right)}. \nonumber
 \end{eqnarray}
where $\log \Lambda^{(k)}_{k,bulk}(x,N)$ is the term proportional to $N$.
By taking the $r \to \infty$ limit of the equations of the critical RSOS($k,q$) lattice model, we obtain
eqs. (\ref{119}) with the $\mathcal{D}(x) \to \hat{\mathcal{D}}_N (x)$ change of the source function, which describe the
$N$ dependence of largest eigenvalue of the transfer matrix $T_{k}(x)$ of the isotropical spin-$k/2$ vertex model.
The equations (\ref{121}) with the $\mathcal{D}(x) \to \hat{\mathcal{D}}_N (x)$ change are though to describe 
the the finite size effects in the anisotropic higher spin vertex model at the spin value $S=k/2$.
The $1/N$ corrections to the largest eigenvalue of the transfer matrices can be analytically calculated
and in every case the well-known results were obtained.


\section{Summary and Conclusions}

From the inhomogeneous RSOS($k,q$) lattice model considered on the critical line separating
regime III and IV, we derived two different sets of 
nonlinear integral equations describing the finite-size effects in the lattice model. The appropriate 
continuum limit of the equations describe the finite-size
dependence of the ground state energy of the massive $\phi_{\mbox{id,id,adj}}$ perturbation of the coset
$SU(2)_{k} \times SU(2)_{r-k-2}/SU(2)_{r-2}$. 
These two sets of nonlinear integral equations are based on the two different formulation of the same model.
One set of equations is derived in the RSOS($k,q$) model, and the other set of equations is derived in the 
RSOS($r-k-2,r-q-2$) model, using their equivalence. These two formulations correspond to two different
representation of the solution of the same (apart from a trivial gauge transformation) T-system.
 These two different representations of the solutions
are characterized by the solutions of two different Bethe Ansatz equations.

Then we glued these two sets of nonlinear integral equations together and obtained two component nonlinear
integral equations for the finite size effects. Taking the $r \to \infty$ limit of our equations we obtained two component 
nonlinear integral equations
corresponding to the case of the isotropic six-vertex model, and the continuum limit of which describe the finite-size effects
in the current-current perturbation of the $SU(2)_k$ WZW model.
Modifying the kernel functions of these equations we proposed two component nonlinear integral equations for the
fractional supersymmetric sine-Gordon models. The lattice version of these equations are thought to describe 
the finite-size effects in the anisotropic higher spin vertex model.

We performed analytical and numerical tests of our equations and all the results of these 
tests made us confident that our equations are correct. The advantage of our equations is that the number of unknown functions are
minimal, and in the case of the FSSG models the equations can be defined for all real values of the coupling constant.
Another advance of having such two component nonlinear integral equations appears in the investigation
of the excited state problem of the models, where only two components have to be treated, instead of the
large number of components of the TBA equations. 
The treatment of excited states in our models is left for future investigation.

\vspace{1cm}
{\tt Acknowledgments}

\noindent 
I would like to thank J\'anos Balog for useful discussions and Zolt\'an Bajnok and G\'abor Tak\'acs 
for the critical reading of the manuscript.
The author acknowledges the financial support provided through
the European Community's Human Potential Programme under contract HPRN-CT-2002-00325, 'EUCLID'. 
This investigation was also supported in part by the 
Hungarian National Science Fund OTKA (under T043159 and T049495) and by INFN Grant TO12.

\end{document}